\documentclass[conference,onecolumn]{IEEEtran}

\IEEEoverridecommandlockouts

\usepackage{cite}
\usepackage{amsmath,amssymb}
\usepackage{graphicx}
\usepackage{booktabs}
\usepackage{array}
\usepackage{xcolor}
\usepackage{listings}
\usepackage{url}
\usepackage{hyperref}
\usepackage{microtype}
\usepackage{tikz}
\usepackage{placeins}
\usepackage{fancyhdr}
\usepackage{xurl}
\usetikzlibrary{arrows.meta,positioning,fit,calc}
\hypersetup{hidelinks}

\newcommand{\mcode}[1]{\text{\texttt{#1}}}

\definecolor{opBlue}{HTML}{3B82F6}
\definecolor{opBlueFill}{HTML}{EFF6FF}
\definecolor{opBlueText}{HTML}{1E40AF}
\definecolor{opGreen}{HTML}{22C55E}
\definecolor{opGreenFill}{HTML}{F0FDF4}
\definecolor{opGreenText}{HTML}{15803D}
\definecolor{opAmber}{HTML}{FCD34D}
\definecolor{opAmberFill}{HTML}{FEF3C7}
\definecolor{opAmberText}{HTML}{92400E}
\definecolor{opRed}{HTML}{EF4444}
\definecolor{opRedFill}{HTML}{FEF2F2}
\definecolor{opRedText}{HTML}{DC2626}
\definecolor{opPurple}{HTML}{8B5CF6}
\definecolor{opPurpleFill}{HTML}{F5F3FF}
\definecolor{opPurpleText}{HTML}{7C3AED}

\newenvironment{floatbox}{\begin{minipage}{0.92\linewidth}\centering}{\end{minipage}}

\lstdefinelanguage{json}{
  morestring=[b]",
  showstringspaces=false,
  morecomment=[l]{//},
  morecomment=[s]{/*}{*/},
  keywords={true,false,null},
  sensitive=false
}

\title{OpenPort Protocol: A Security Governance Specification for AI Agent Tool Access}

\author{%
\parbox[t]{0.48\linewidth}{\centering
  Genliang Zhu\\
  Accentrust\\
  Georgia Institute of Technology
}
\hfill
\parbox[t]{0.48\linewidth}{\centering
  Chu Wang\\
  Accentrust\\
  University of Illinois Urbana-Champaign
}
\\[0.8em]
\parbox[t]{0.32\linewidth}{\centering
  Ziyuan Wang\\
  Duke University
}
\hfill
\parbox[t]{0.32\linewidth}{\centering
  Zhida Li\\
  New York Institute of Technology-Vancouver
}
\hfill
\parbox[t]{0.32\linewidth}{\centering
  Qiang Li\\
  Georgia Institute of Technology
}
\thanks{Stewardship: Accentrust and the OpenPort Protocol authors.}%
\thanks{Correspondence: \href{mailto:research@accentrust.com}{research@accentrust.com}.}%
}

\begin{document}
\maketitle
\pagestyle{fancy}
\fancyhf{}
\renewcommand{\headrulewidth}{0pt}
\renewcommand{\footrulewidth}{0pt}
\fancyfoot[R]{\thepage}
\thispagestyle{fancy}

\begin{abstract}
AI agents increasingly require direct, structured access to application data and actions, but production deployments still
struggle to express and verify the governance properties that matter in practice: least-privilege authorization,
controlled write execution, predictable failure handling, abuse resistance, and auditability.
This paper introduces OpenPort Protocol (OPP), a governance-first specification for exposing application tools through a secure
server-side gateway that is model- and runtime-neutral and can bind to existing tool ecosystems.
OpenPort defines authorization-dependent discovery, stable response envelopes with machine-actionable \texttt{agent.*}
reason codes, and an authorization model combining integration credentials, scoped permissions, and ABAC-style policy
constraints.
For write operations, OpenPort specifies a risk-gated lifecycle that defaults to draft creation and human review, supports
time-bounded auto-execution under explicit policy, and enforces high-risk safeguards including preflight impact binding and
idempotency.
To address time-of-check/time-of-use drift in delayed approval flows, OpenPort also specifies an optional State Witness
profile that revalidates execution-time preconditions and fails closed on state mismatch.
Operationally, the protocol requires admission control (rate limits/quotas) with stable 429 semantics and structured audit
events across allow/deny/fail paths so that client recovery and incident analysis are deterministic.
We present a reference runtime and an executable governance toolchain (layered conformance profiles, negative security
tests, fuzz/abuse regression, and release-gate scans) and evaluate the core profile at a pinned release tag using
artifact-based, externally reproducible validation.
\end{abstract}


\section{Introduction}
Modern AI agents operate by calling tools: structured functions or actions that read data and perform controlled writes
in external systems.
While early deployments rely on browser automation or one-off internal APIs, production environments require clear and
verifiable guarantees: least-privilege access~\cite{saltzer1975}, tenant isolation, revocation, operational rate limits,
and a complete audit trail.

OpenPort Protocol (OPP) targets the gap between ``agents can call tools'' and ``tools can be safely exposed in real systems.''
The core observation is that tool discovery is not the hard part; authorization and governance are.
A safe agent access layer must define the authorization surface, encode risk constraints for write operations, and make
every decision auditable and operable.
In other words, the ``killer test'' for production tool exposure is not a tool-list demo, but real-world authorization
flows and predictable rate-limit behavior under failure and abuse.

\textbf{Contributions.} This paper makes three contributions:
\begin{enumerate}
  \item A protocol-level specification for secure tool exposure to AI agents, including endpoints, response envelopes, tool metadata, and error taxonomy.
  \item A security governance model that combines scopes and policy constraints with a draft-first write pipeline, high-risk safeguards (preflight, idempotency, and an optional state-witness profile for TOCTOU), mandatory audit events, and rate limits/quotas.
  \item An engineering methodology for safe open-source evolution via conformance tests, fuzz/abuse regression, and release-gate checks that operationalize protocol invariants.
\end{enumerate}

\section{Problem Statement and Threat Model}
OpenPort focuses on the security and governance gap between ``tool exposure'' and ``safe tool exposure.''
While tool discovery and structured schemas are necessary for agents to interact with applications, they are not
sufficient for production deployments.
In practice, AI agents are probabilistic callers: they may call the wrong tool, call the right tool with the wrong
arguments, retry unexpectedly, or be influenced by untrusted inputs (for example, prompt injection) to perform unsafe actions.
Treating agent tool access as ``just another API client'' fails to capture these operational failure modes.
This paper therefore frames the problem as a governance problem: how to expose application data and actions to agents
through a narrow, explicitly authorized, rate-limited, and auditable interface that is safe under both malicious abuse
and benign agent mistakes.

\subsection{System Model and Trust Boundaries}
OpenPort assumes a server-side gateway that mediates all agent access to an application.
The gateway is the enforcement point responsible for authentication, authorization, rate limiting, audit emission, and
write governance.
Domain-specific logic and identity models remain behind adapters, allowing products to preserve internal schemas while
still adopting a standardized governance surface.
We assume the application operator controls the gateway and the admin control plane, and that all external access is
over TLS.

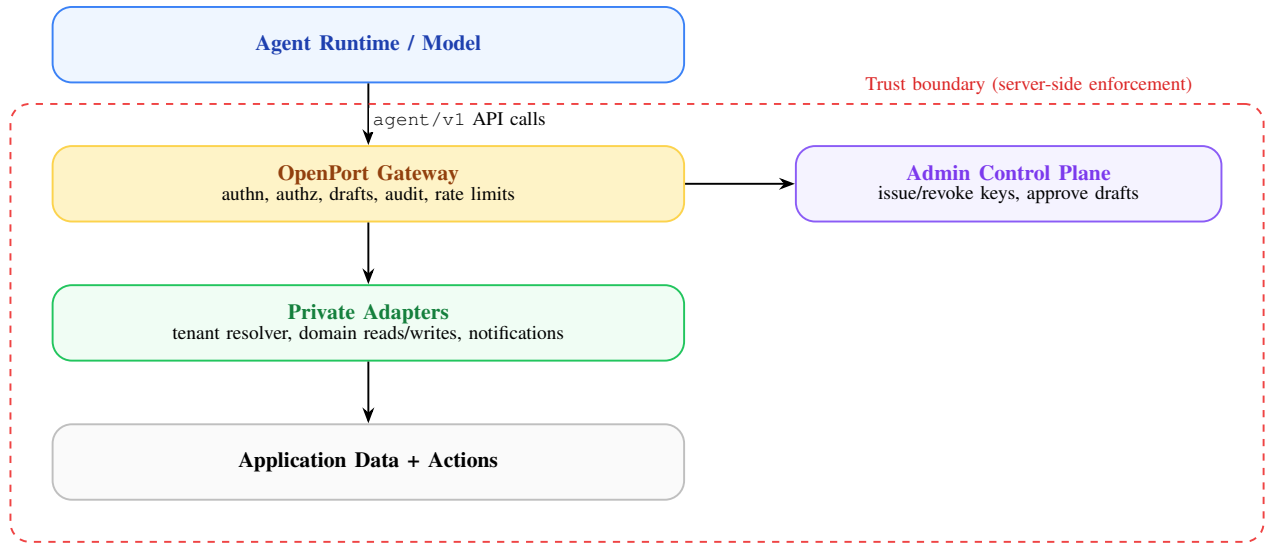
\begin{figure}[t]
\centering
\begin{floatbox}
\resizebox{\linewidth}{!}{%
\begin{tikzpicture}[
  >=Stealth,
  box/.style={draw, rounded corners=2.8mm, thick, align=center, minimum width=9.2cm, minimum height=1.1cm, inner sep=6pt, font=\small},
  side/.style={draw, rounded corners=2.8mm, thick, align=center, minimum width=6.2cm, minimum height=1.1cm, inner sep=6pt, font=\small},
  edge/.style={->, thick},
  elabel/.style={font=\footnotesize, align=center, fill=white, inner sep=1.5pt}
]
\node[box, draw=opBlue, fill=opBlueFill] (agent) {\textcolor{opBlueText}{\textbf{Agent Runtime / Model}}};
\node[box, draw=opAmber, fill=opAmberFill, below=9mm of agent] (gateway) {\textcolor{opAmberText}{\textbf{OpenPort Gateway}}\\[-2pt]\footnotesize authn, authz, drafts, audit, rate limits};
\node[box, draw=opGreen, fill=opGreenFill, below=9mm of gateway] (adapters) {\textcolor{opGreenText}{\textbf{Private Adapters}}\\[-2pt]\footnotesize tenant resolver, domain reads/writes, notifications};
\node[box, draw=black!25, fill=black!2, below=9mm of adapters] (data) {\textbf{Application Data + Actions}};
\node[side, draw=opPurple, fill=opPurpleFill, right=16mm of gateway] (admin) {\textcolor{opPurpleText}{\textbf{Admin Control Plane}}\\[-2pt]\footnotesize issue/revoke keys, approve drafts};

\draw[edge] (agent) -- node[elabel, right, pos=0.60] {\texttt{agent/v1} API calls} (gateway);
\draw[edge] (gateway) -- (adapters);
\draw[edge] (adapters) -- (data);
\draw[edge] (gateway.east) -- (admin.west);

% Trust boundary: server-side operator-controlled components.
\node[draw=opRed, dashed, rounded corners=2.8mm, thick, fit=(gateway) (adapters) (data) (admin), inner sep=6mm,
  label={[font=\footnotesize, text=opRedText]above right:Trust boundary (server-side enforcement)}] {};
\end{tikzpicture}%
}
\end{floatbox}
\caption{System model and trust boundaries.}
\label{fig:trust-model}
\end{figure}

\subsection{Assets}
OpenPort treats the following as protected assets:
\begin{itemize}
  \item \textbf{Agent credentials}: agent keys and delegated tokens, including their metadata (key identifiers,
  issuance time, revocation state).
  \item \textbf{Tenant/workspace data}: the confidentiality of records reachable through read tools and the integrity of
  state modified by write tools.
  \item \textbf{Write intent and effects}: the correctness of drafts and executions, including protection against
  replay, duplicates, and unintended side effects.
  \item \textbf{Audit stream}: decision records and execution traces required for incident response and compliance.
  \item \textbf{Operational budget}: rate-limit and quota capacity, including protection against cost amplification
  (expensive queries, bulk exports, and high-frequency polling).
\end{itemize}

\subsection{Assumptions}
The threat model is intentionally pragmatic and matches typical production constraints:
\begin{itemize}
  \item Transport is protected by TLS and the gateway is the single externally reachable enforcement point.
  \item Private adapters and application databases are not directly reachable by the agent runtime.
  \item Attackers may observe agent-visible content (tool descriptions, responses, and error messages) and may cause the
  agent to retry or to call unintended tools.
  \item Credentials may leak; therefore, OpenPort assumes that revocation, rate limiting, and auditability are required
  even for valid-looking requests.
\end{itemize}

\subsection{Adversary Model}
We consider both malicious and accidental failures that are characteristic of agent tool access:
\begin{itemize}
  \item \textbf{External attacker}: attempts to steal tokens, enumerate tools, or exploit weak authorization checks.
  \item \textbf{Compromised agent runtime}: uses valid credentials to maximize impact (abuse, exfiltration).
  \item \textbf{Prompt injection / untrusted content}: influences an agent to request unsafe actions.
  \item \textbf{Benign failure}: retries, partial failures, and non-idempotent writes causing duplicate effects.
  \item \textbf{Operator mistakes}: overbroad scopes or misconfigured policies that silently increase blast radius.
\end{itemize}
We do not assume the model itself is trustworthy; instead, OpenPort constrains the \emph{effects} of tool calls by
requiring server-side enforcement and reviewable write paths.

\subsection{Key Threats}
When an application is opened to AI agents, a recurring set of production risks emerges:
\begin{enumerate}
  \item \textbf{Token leakage} through prompts, logs, repositories, or CI systems.
  \item \textbf{Tenant/workspace boundary bypass} due to client-supplied identifiers or inconsistent checks.
  \item \textbf{Tool enumeration and capability leakage} where manifests, schemas, or error details reveal privileged
  capabilities or sensitive fields.
  \item \textbf{Unapproved destructive writes} (delete/export/irreversible actions) caused by automation mistakes or prompt injection.
  \item \textbf{Replay and duplicate effects} caused by retries, race conditions, or non-idempotent execution.
  \item \textbf{Insufficient auditability} where decisions and side effects cannot be reconstructed.
  \item \textbf{Abuse and overload} via excessive calls (DoS), expensive operations, or cost amplification.
  \item \textbf{Sensitive data overexposure} through error details and logs that inadvertently capture payloads.
\end{enumerate}

\subsection{Security Goals}
OpenPort Protocol addresses these threats by specifying server-side governance as protocol requirements.
Rather than leaving security decisions as ``implementation-defined,'' OpenPort promotes a small set of verifiable
properties that can be tested in conformance suites and enforced in release gates.
We state these properties as concrete security invariants:
\begin{enumerate}
  \item \textbf{Deny-by-default authorization.} A request MUST be denied unless explicitly allowed by server-side scope
  grants and policy constraints; there is no implicit access.
  \item \textbf{Server-enforced tenant/workspace boundaries.} Every read and write MUST be bound to a server-verified
  tenant/workspace context; client-supplied identifiers MUST NOT be trusted as the boundary.
  \item \textbf{Authorization-dependent discovery.} \texttt{/manifest} MUST reflect current authorization: tools,
  fields, and capabilities that are not permitted MUST NOT be exposed via discovery responses.
  \item \textbf{Immediate revocation.} Revoking a key or app MUST take effect immediately on all endpoints; denials due
  to revocation MUST be audited with a stable reason code.
  \item \textbf{Draft-first writes.} Write actions MUST default to creating drafts; direct execution is only permitted
  when explicitly enabled by policy, and SHOULD be time-bounded.
  \item \textbf{High-risk execution safeguards.} High-risk writes SHOULD require step-up confirmation and MUST support
  preflight impact hashing and idempotency keys to reduce unintended effects.
  \item \textbf{Replay and duplicate protection.} For write intents, the pair \{\texttt{appId}, \texttt{idempotencyKey}\}
  MUST uniquely identify an execution outcome and MUST be safely de-duplicated.
  \item \textbf{Operable abuse controls.} Implementations MUST enforce baseline rate limits and return stable 429
  semantics; servers SHOULD include backoff guidance (for example, \texttt{Retry-After}) and SHOULD support quotas by tool and
  tenant.
  \item \textbf{Audit-first governance.} Allow/deny/fail decisions and execution outcomes MUST emit structured audit
  events with stable reason codes and incident-response metadata (for example, key/app identifiers and request metadata).
  \item \textbf{Data minimization.} Logs, error details, and audit payloads SHOULD redact sensitive fields and MUST NOT
  contain secrets such as bearer tokens.
\end{enumerate}

\begin{table}[t]
\caption{Threats and required controls in OpenPort.}
\label{tab:threats-controls}
\centering
\begin{floatbox}
\begin{tabular}{p{0.36\linewidth} p{0.58\linewidth}}
\toprule
\textbf{Threat} & \textbf{Required protocol controls} \\
\midrule
Token leakage & Revocation, short-lived execution windows, rate limits, audit on every decision \\
Cross-tenant access & Server-side tenant resolver; deny-by-default; policy constraints; no client-trusted tenant IDs \\
Tool enumeration & AuthZ-dependent manifest; schema and error redaction; deny-by-default exposure \\
Destructive write misuse & Draft-first default; step-up confirmation; human approval chain; explicit risk tiers \\
Replay / duplicates & Idempotency keys; execution de-duplication per app and key \\
Audit gaps & Structured allow/deny/fail audit events with stable reason codes and request metadata \\
Abuse / DoS & Per-key/per-IP rate limiting; quotas by tool; predictable 429 behavior \\
Sensitive logging & Redaction rules; separation of operational logs vs immutable audit stream \\
\bottomrule
\end{tabular}
\end{floatbox}
\end{table}

\subsection{Non-goals}
OpenPort does not replace product identity systems, and it does not guarantee safety against a malicious administrator
who controls the gateway and the admin plane.
It also does not attempt to solve general prompt safety; instead, it constrains the \emph{effects} of tool calls via
governance requirements (least privilege, reviewable writes, and audited decisions).

\FloatBarrier
\section{Protocol Overview}
\subsection{Design Principles}
OpenPort is designed as a governance-first protocol surface: it standardizes the controls that must hold when exposing
application data and actions to probabilistic callers (agent runtimes) rather than assuming a fully correct client.
Its core principles are: (i) authorization-dependent, machine-readable tool discovery; (ii) server-side enforcement of
authorization, policy constraints, and risk controls; (iii) stable, parseable response envelopes and \texttt{agent.*}
reason codes so failures are operable; and (iv) a narrow write interface that defaults to reviewable drafts and can be
strengthened with high-risk safeguards (preflight hashing, idempotency, and optional state-witness preconditions).
Operationally, OpenPort treats admission control and auditability as first-class requirements: requests must be rate
limited/quota controlled and every allow/deny/fail path must be reconstructable from structured audit events.

\begin{figure}[t]
\centering
\begin{floatbox}
\begin{tikzpicture}[
  >=Stealth,
  box/.style={draw, rounded corners=2.8mm, thick, align=center, minimum width=0.78\linewidth, minimum height=1.10cm, inner sep=5pt, font=\small},
  edge/.style={->, thick},
  elabel/.style={font=\footnotesize, align=center, fill=white, inner sep=1.5pt}
]
\node[box, draw=opBlue, fill=opBlueFill] (ecos) {\textcolor{opBlueText}{\textbf{Tool Ecosystems / Runtimes}}\\[-2pt]\footnotesize MCP clients\quad$\bullet$\quad WebMCP pages\quad$\bullet$\quad custom SDKs};
\node[box, draw=opAmber, fill=opAmberFill, below=7mm of ecos] (core) {\textcolor{opAmberText}{\textbf{OpenPort Protocol (Core)}}\\[-2pt]\footnotesize authZ-dependent discovery\quad$\bullet$\quad stable envelopes\\[-2pt]\footnotesize draft-first writes\quad$\bullet$\quad rate limits\quad$\bullet$\quad audit};
\node[box, draw=opGreen, fill=opGreenFill, below=7mm of core] (deploy) {\textcolor{opGreenText}{\textbf{OpenPort Deployment}}\\[-2pt]\footnotesize gateway\quad$\bullet$\quad admin plane\quad$\bullet$\quad adapters};

\draw[edge] (ecos) -- node[elabel, right, xshift=6pt, text=opPurpleText, pos=0.52] {bindings (optional)} (core);
\draw[edge] (core) -- node[elabel, right, xshift=6pt, pos=0.52] {server-side enforcement} (deploy);
\end{tikzpicture}
\end{floatbox}
\caption{OpenPort separates governance semantics from tool ecosystems: compatibility is achieved via optional bindings.}
\label{fig:binding-overview}
\end{figure}
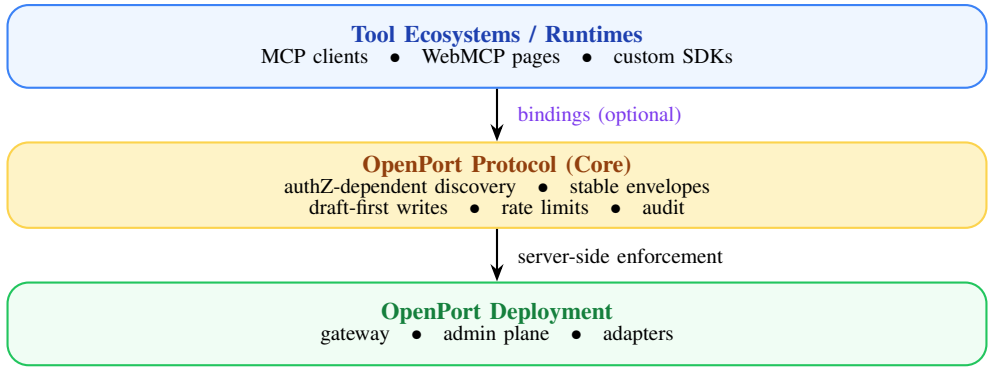

The protocol is domain-extensible: products implement domain reads and actions via adapters, while OpenPort fixes the
governance semantics that must remain consistent across domains.
Figure~\ref{fig:binding-overview} summarizes this separation: OpenPort specifies governance semantics and a stable
server-side interface, while compatibility with existing tool ecosystems is achieved through optional bindings rather
than by delegating enforcement to each ecosystem.
Bindings are informative and are evaluated separately from the core conformance claims in this paper.

\subsection{Versioning and Compatibility}
OpenPort uses explicit versioning in its URL path (for example, \texttt{agent/v1}) to support long-lived clients.
Within a major version, servers SHOULD evolve the protocol in a backward-compatible manner (additive fields, additive
tools, and stricter governance defaults that remain safe for existing clients).
Clients SHOULD ignore unknown fields and treat tool discovery as dynamic, because permitted tools and schemas may change
with scopes, policy constraints, and incident response actions such as revocation.

\subsection{Core Objects}
OpenPort standardizes a small set of objects used by both the protocol and governance layer:
\begin{itemize}
  \item \textbf{Integration App}: a managed integration configuration with scopes and policy.
  \item \textbf{Agent Key}: a revocable token bound to an app (stored as a hash server-side).
  \item \textbf{Tool}: a machine-readable description of an action or read operation, including schemas and governance
  metadata.
  \item \textbf{Draft}: a reviewable representation of a requested write action.
  \item \textbf{Execution}: the outcome of executing a draft or an allowed write action.
  \item \textbf{Audit Event}: an immutable record of allow/deny/fail decisions and executions.
\end{itemize}

\subsection{Tool Manifest}
OpenPort treats tool discovery as an authorization-sensitive API.
The agent runtime fetches \texttt{GET /api/agent/v1/manifest} to obtain the integration identity and a list of tools
permitted under the presented credential.
Each tool includes governance metadata (required scopes, risk tier, and confirmation requirements) and machine-readable
schemas for inputs and outputs.
The manifest MAY include HTTP hints (method and path) to support runtimes that want to provide clickable traces or
debugging output, but correctness MUST NOT depend on clients calling arbitrary paths.
Because the manifest is agent-visible, it MUST avoid capability leakage: tools and fields that are not authorized for
the current integration MUST NOT appear, and error details SHOULD be redacted to avoid revealing privileged names.

Formally, let $\mathcal{T}$ be the global set of tools, let $\mathrm{ReqScopes}(t)$ be the required scope set for tool
$t$, and let $\mathrm{Scopes}(a)$ be the scopes granted to integration app $a$.
Let $\mathrm{PolicyAllows}(a, t)$ denote a predicate that captures additional policy constraints on exposure (for example, field
redaction and disabled capabilities). OpenPort defines the visible tool set as:
\begin{equation}
\begin{aligned}
\mathrm{VisibleTools}(a) = \{ t \in \mathcal{T} \mid {}&\mathrm{ReqScopes}(t) \subseteq \mathrm{Scopes}(a)\\
&\wedge\ \mathrm{PolicyAllows}(a, t) \}.
\end{aligned}
\label{eq:visible-tools}
\end{equation}
This equation expresses the core ``authorization-dependent discovery'' requirement: discovery output is a function of
current authorization state, not a static listing.

OpenPort uses a small, opinionated risk scale to support consistent governance.
\textbf{Low} risk tools are typically read-only or trivially reversible operations; \textbf{medium} risk tools perform
write operations that are expected to be reversible or bounded; and \textbf{high} risk tools include destructive writes,
bulk exports, or other actions with significant blast radius.
Risk tier is part of tool metadata to support both server-side enforcement (draft-first defaults and high-risk
safeguards) and client-side UX (when to request preflight, how to explain required approvals).

\begin{lstlisting}[language=json,caption={Conceptual tool object in OpenPort manifest.}]
{
  "name": "transaction.hard_delete",
  "description": "Permanently delete a transaction.",
  "requiredScopes": ["transaction.delete"],
  "risk": "high",
  "requiresConfirmation": true,
  "http": { "method": "POST", "path": "/api/agent/v1/actions" },
  "inputSchema": { "type": "object", "properties": { "payload": { "type": "object" } } },
  "outputSchema": { "type": "object", "properties": { "deleted": { "type": "object" } } }
}
\end{lstlisting}

\subsection{Core Endpoints}
OpenPort defines a minimal agent-facing protocol surface:
\begin{lstlisting}[language=bash,caption={OpenPort Protocol agent/v1 endpoints (conceptual).}]
GET  /api/agent/v1/manifest
GET  /api/agent/v1/ledgers
GET  /api/agent/v1/transactions
POST /api/agent/v1/preflight
POST /api/agent/v1/actions
GET  /api/agent/v1/drafts/{id}
\end{lstlisting}

Read endpoints (for example, \texttt{/ledgers}, \texttt{/transactions}) are illustrative; implementations MAY expose additional
domain reads as long as they follow the stable envelope, enforce the same authorization model, and emit audit events.
In contrast, writes are intentionally funneled through a small number of governance-aware endpoints.

An admin control plane is required for safe operation and is described below.

\subsection{Domain Extensibility and Adapters}
OpenPort intentionally separates \emph{governance semantics} from \emph{domain semantics}.
Domain adapters implement product-specific reads and actions (for example, accounting, CRM, ticketing) and translate them into
the OpenPort tool model.
This boundary has two protocol consequences:
\begin{itemize}
  \item \textbf{Extensible reads.} Implementations MAY add new read endpoints, but each MUST be declared (directly or
  indirectly) through authorization-dependent discovery and MUST enforce the same scope/policy/tenant controls as core
  endpoints.
  \item \textbf{Centralized writes.} Implementations SHOULD route writes through a small number of governance-aware
  endpoints (for example, \texttt{/preflight} and \texttt{/actions}) so that draft-first, step-up, idempotency, and audit
  behavior remains consistent across all write tools.
\end{itemize}
In the reference runtime, this takes the form of a tool registry that binds each tool name to an adapter-backed
function and attaches governance metadata (required scopes, risk tier, and optional impact computation).

\subsection{Admin Control Plane (Required)}
OpenPort relies on a human-managed admin control plane to close the authorization loop.
While product identity is out of scope, production deployments require a minimal set of governance actions:
\begin{itemize}
  \item create and revoke integration apps and keys (issuance and rotation)
  \item update policy constraints and auto-execute windows
  \item list, approve, and reject drafts
  \item export audit events and operational summaries
\end{itemize}
This admin plane is also the natural place to enforce organizational procedures (for example, separation of duties) and to
integrate with incident response workflows.

\subsection{Write Lifecycle: Preflight, Drafts, Execution}
OpenPort separates three phases of write operations:
\begin{itemize}
  \item \textbf{Preflight} (\texttt{POST /preflight}) optionally computes an impact summary and returns an impact hash that
  can be bound to later execution. Under an optional stronger governance profile, preflight MAY also return a
  \texttt{stateWitnessHash} that binds execution to a server-observed resource version/state. Implementations MAY also
  return a short-lived preflight handle (\texttt{preflightId})
  that allows clients to reuse the exact payload and hash without resending or regenerating content.
  High-risk policies may require preflight.
  \item \textbf{Request} (\texttt{POST /actions}) submits a tool name and payload, and produces either a draft (default)
  or an execution result (when explicitly permitted).
  \item \textbf{Review/Execute} occurs via the admin plane for draft approval, followed by execution under policy
  constraints (idempotency, preflight hash binding, and step-up requirements).
\end{itemize}
Drafts expose a stable status model (for example, draft, confirmed, canceled, failed) and are queryable via
\texttt{GET /drafts/\{id\}}. This allows agent runtimes to treat writes as asynchronous and reviewable by design.

\subsubsection{Preflight Hash Binding}
Preflight binds execution to a computed impact summary by returning a stable digest over canonical inputs.
Let $H(\cdot)$ be a cryptographic hash function and let $\mathcal{C}(\cdot)$ be a deterministic canonical encoding
(for the reference profile, RFC 8785 JSON Canonicalization Scheme (JCS) over JSON values~\cite{rfc8785}).
Let $I(t, x)$ be the server-side impact function for tool $t$ and payload $x$ (possibly adapter-backed).
OpenPort defines the preflight hash as:
\begin{equation}
h = H\big(\mathcal{C}(t, x, I(t, x))\big).
\label{eq:preflight-hash}
\end{equation}
If a policy requires preflight for a request, the server MUST deny execution unless the client supplies a preflight hash
that matches the server-computed value in Eq.~\eqref{eq:preflight-hash}.
To make this binding robust under agent non-determinism, clients SHOULD treat \texttt{/preflight} and \texttt{/actions}
as one write intent: do not generate a second payload after preflight. A server MAY provide \texttt{preflightId} as a
convenience handle that resolves to the cached payload and hash. Such handles SHOULD be TTL-bound and scoped to the
credential context (for example, app/key/actor) that created them. If a \texttt{preflightId} is provided but cannot be resolved
(expired, unknown, or wrong credential context), the server SHOULD fail closed with \url{agent.preflight_not_found}.

\subsubsection{State Witness Preconditions (Optional Strong Profile)}
Preflight hash binding protects against payload non-determinism and intent swapping, but it does not fully address
time-of-check to time-of-use (TOCTOU) risks when human approvals are delayed and the underlying world state can change.
OpenPort therefore defines an optional stronger governance profile: \emph{State Witness / Preconditions}.
Implementations MAY support this profile. Implementations that claim this profile MUST enforce the preconditions described
below when a request or draft is bound to a state witness hash.

Let $W(t, x)$ be a server-side witness function that returns a minimal, non-secret snapshot of the relevant resource
state for tool $t$ and payload $x$ (for example, an HTTP \texttt{ETag}-like value, a \texttt{resourceVersion}-like field, or an
adapter-defined version tuple). OpenPort defines a witness hash:
\begin{equation}
w = H\big(\mathcal{C}(W(t, x))\big).
\label{eq:state-witness-hash}
\end{equation}
If a draft is bound to $w$ (or a client supplies \texttt{stateWitnessHash} directly), the server MUST recompute the
current witness at execution time and MUST fail closed unless the current witness hash matches $w$.
On mismatch, the server MUST deny with a stable reason code \url{agent.precondition_failed} and MUST NOT execute side
effects. Clients SHOULD treat this as a state change: rerun \texttt{/preflight}, refresh operator approval, and avoid
blind retries.
This mechanism is specified as an optional profile extension and is evaluated separately from the pinned
\texttt{v0.1.0} core evaluation claims in Section~\ref{sec:evaluation}.

\subsubsection{Action Request Schema}
Action requests use an explicit tool name and a free-form payload whose structure is governed by the tool's input
schema in the manifest.
To support safe retries and operational tracing, OpenPort defines optional fields for request correlation and replay
protection:
\begin{lstlisting}[language=json,caption={Conceptual OpenPort action request.}]
{
  "action": "transaction.hard_delete",
  "payload": { "...": "..." },
  "preflightId": "pfl_...",
  "execute": true,
  "forceDraft": false,
  "requestId": "req_...",
  "idempotencyKey": "idem_...",
  "justification": "operator intent for high-risk execution",
  "preflightHash": "sha256(...preflight impact...)",
  "stateWitnessHash": "sha256(...witness...)"
}
\end{lstlisting}
When \texttt{preflightId} is provided, servers MAY reuse a cached payload and preflight hash, allowing clients to omit
\texttt{payload} as long as the resolved payload matches the requested action and credential scope.
The server MUST decide whether execution is permitted. If execution is not permitted, the server MUST return a draft
and MUST provide a machine-readable denial reason code for the auto-execute request.
If an idempotency key is provided and a matching execution already exists, the server SHOULD return the prior execution
outcome rather than re-executing the tool, to make agent retries safe.

\subsubsection{Auto-Execute Eligibility Predicate}
OpenPort treats auto-execution as an explicit, time-bounded capability.
Let $a$ be an integration app, let $t$ be a tool, and let $\mathrm{cfg}(a)$ be the auto-execute configuration state.
Let $\mathrm{exp}(a,t)$ denote the configured expiration timestamp for the relevant auto-execute window and let
$\mathrm{AllowList}(a,t)$ denote an optional allowlist of tool names.
Let $\mathrm{Req}(r)$ be the action request fields (execute, forceDraft, justification, idempotencyKey, and either
\texttt{preflightHash} or a \texttt{preflightId} that resolves to one, and optionally \texttt{stateWitnessHash} under
the State Witness profile).
OpenPort defines an eligibility predicate:
\begin{equation}
\begin{aligned}
\mathrm{AutoExecAllowed}(r) =\ &\mathrm{Req}(r).\mathrm{execute} \wedge \neg \mathrm{Req}(r).\mathrm{forceDraft}\\
&\wedge\ \mathrm{Enabled}(\mathrm{cfg}(a), t) \wedge (\mathrm{ts} < \mathrm{exp}(a,t))\\
&\wedge\ \Big(\mathrm{AllowList}(a,t)=\emptyset\ \vee\\
&\qquad t \in \mathrm{AllowList}(a,t)\Big)\\
&\wedge\ \Big(\mathrm{risk}(t)\neq \mcode{high}\ \vee\\
&\qquad\big(\mathrm{Req}(r).\mathrm{justification}\neq \emptyset\ \wedge\\
&\qquad\mathrm{HighRiskGuards}(r)\big)\Big),
\end{aligned}
\label{eq:autoexec}
\end{equation}
where $\mathrm{HighRiskGuards}(r)$ includes idempotency and preflight requirements when configured, and preflight hash
matching uses Eq.~\eqref{eq:preflight-hash}.
If $\mathrm{AutoExecAllowed}(r)$ is false, the server MUST return a draft and MUST return a stable denial reason code.

\subsubsection{Idempotency as a Deterministic Execution Map}
Idempotency is expressed as a deterministic mapping from an integration app and idempotency key to an execution outcome.
Let $a$ be an integration app identifier and let $k$ be an idempotency key. OpenPort defines:
\begin{equation}
E(a, k) \to \mcode{Execution},
\label{eq:idempotency-map}
\end{equation}
with the invariant that if $E(a, k)$ is already defined, then subsequent requests carrying the same $(a, k)$ MUST NOT
re-execute side effects and SHOULD return the existing execution record.
This invariant converts agent retries and partial failures into safe, replayable outcomes.

\subsubsection{Policy Snapshotting}
To make approvals and incident response tractable, OpenPort recommends snapshotting the governance-relevant policy
inputs at the time a draft is created (for example, required scopes, risk tier, and auto-execute configuration).
This supports later review of ``what the system believed was permitted'' at the time of the request, even if policies
change before approval.

\subsection{Response Envelope and Error Taxonomy}
Every endpoint returns a stable response envelope:
\begin{itemize}
  \item success: \texttt{\{ ok: true, code, data \}}
  \item error: \texttt{\{ ok: false, code, message, details? \}}
\end{itemize}
This stability is a core requirement for agent runtimes that need predictable parsing and recovery strategies.
The \texttt{code} field is a stable machine-facing identifier. For operational controls such as rate limiting, servers
SHOULD return 429 with backoff guidance (for example, \texttt{Retry-After})~\cite{rfc9110,rfc6585}. For authorization denials,
servers SHOULD return stable denial codes to support safe retry and operator escalation.
OpenPort uses a small error taxonomy to make denial behavior predictable (invalid/expired token, scope or policy denial,
unknown/invalid action, preflight required/mismatch/not found, idempotency required/replay, and rate limited).
Optional stronger profiles MAY add additional stable denial codes; for example, the State Witness profile introduces
\url{agent.precondition_failed} when an execute-time precondition does not hold.

\section{Authorization Model}
\subsection{Token Types}
OpenPort supports two authorization modes that cover most production deployments:
\begin{enumerate}
  \item \textbf{Integration tokens} for machine-to-machine access, bound to an integration app/key and constrained by
  server-side scopes and policy constraints.
  \item \textbf{Delegated tokens} for ``act on behalf of'' flows (for example, OAuth 2.0), which MUST be mapped to minimal
  OpenPort scopes and policy windows before any tool access is granted.
\end{enumerate}
The protocol does not mandate a specific identity provider; it mandates that authorization decisions are made
server-side and are auditable.

\subsection{Credential Presentation and Validation}
Agent requests present credentials as bearer tokens (for example, \texttt{Authorization: Bearer <token>}).
To minimize leakage impact, OpenPort recommends opaque (reference) tokens that are validated by server-side lookup; if
self-contained tokens are used, they SHOULD be short-lived and MUST support immediate revocation.

Validation MUST check, at minimum:
\begin{itemize}
  \item token presence and parseability
  \item token revocation and expiration
  \item integration app status (active vs revoked/disabled)
  \item network policy constraints (for example, IP allowlists) when configured
  \item baseline abuse controls (rate limits) to bound impact under leakage
\end{itemize}
Failures MUST return stable denial codes (for example, invalid token, expired token, policy denied, rate limited) so agent
runtimes can implement safe retry and operator escalation.

\subsection{Authorization Decision Algorithm}
OpenPort constrains authorization to a deterministic decision function. For each request, servers SHOULD evaluate in a
fixed order:
\begin{enumerate}
  \item \textbf{Authenticate}: validate the credential and resolve the integration app.
  \item \textbf{Apply network policy}: enforce IP allowlists or other network constraints.
  \item \textbf{Rate limit}: apply per-key/per-IP limits to bound abuse.
  \item \textbf{Authorize by scope}: ensure the tool's required scopes are granted to the app.
  \item \textbf{Authorize by policy}: apply ABAC-style constraints to restrict the data domain and output surface.
  \item \textbf{Enforce tenant/workspace boundary}: verify all domain access is within the app's permitted boundary.
  \item \textbf{Audit}: emit allow/deny/fail events for authenticated requests with stable reason codes and
  incident-response metadata.
\end{enumerate}
This ordering is not cosmetic: it is what makes decisions predictable, testable, and operable.

\subsubsection{Formal Authorization Predicate}
Let a request be $r = (\tau, t, x, \mathrm{ip}, \mathrm{ts})$ where $\tau$ is a credential, $t$ is a tool name, $x$ is
the payload, and $\mathrm{ip}$ and $\mathrm{ts}$ are request metadata.
Let $\mathrm{App}(\tau)$ resolve the integration app and key for a valid credential.
OpenPort defines authorization as the conjunction of server-side predicates:
\begin{equation}
\begin{aligned}
\mathrm{Allow}(r) =\ &\mathrm{Authn}(\tau)\ \wedge\ \mathrm{Net}(\mathrm{ip}, \mathrm{App}(\tau))\ \wedge\ \mathrm{RL}(r)\\
&\wedge\ \mathrm{ScopeOK}(t, \mathrm{App}(\tau))\ \wedge\ \mathrm{PolicyOK}(t, x, \mathrm{App}(\tau))\\
&\wedge\ \mathrm{BoundaryOK}(t, x, \mathrm{App}(\tau)).
\end{aligned}
\label{eq:allow}
\end{equation}
To ensure stable reason codes, implementations can map the \emph{first failing predicate} in the evaluation order to a
denial code. If the ordered predicates are $(p_1,\ldots,p_n)$, the denial index is:
\begin{equation}
i^\ast = \min\{ i \mid p_i(r) = \mcode{false} \},
\label{eq:first-fail}
\end{equation}
and the reason code is a deterministic function $\mathrm{Code}(p_{i^\ast})$.

\subsection{Least Privilege: Scopes}
Scopes are static capability labels required by tools (for example, \texttt{transaction.read}, \texttt{transaction.write}).
Tools declare \texttt{requiredScopes} in the manifest. The server MUST enforce deny-by-default scope checks; absence of
a required scope implies denial.
When multiple scopes are required, the semantics are conjunctive (all required scopes must be present).

\subsection{Least Privilege: Policy Constraints (ABAC)}
Policy constraints restrict the \emph{data domain} and \emph{presentation surface} beyond what scopes can express.
OpenPort models policy as ABAC-style constraints~\cite{nist800162} and treats it as mandatory enforcement state.
Typical constraints include:
\begin{itemize}
  \item \textbf{Network policy}: IP allowlists for machine clients.
  \item \textbf{Data policy}: allowed resource sets (for example, allowed ledger IDs or organization IDs), bounded time windows
  for queries, and field-level redaction toggles for sensitive attributes.
\end{itemize}
Policy MUST be enforced server-side and SHOULD be reflected in discovery responses (for example, hiding tools or fields that
would be denied).

\subsubsection{A Bounded Query Window Constraint}
A common production control is bounding query time windows to limit both data exposure and cost amplification.
Let $d_{\max}$ be the maximum number of days allowed by policy, and let $(s,e)$ be the effective start and end dates for
a request after defaulting missing bounds.
OpenPort requires the constraint:
\begin{equation}
\Delta(s,e) = \frac{e - s}{\mathrm{day}} \le d_{\max},
\label{eq:max-days}
\end{equation}
otherwise the server MUST deny with a policy reason code.

\subsubsection{Policy Effects on Output}
Policy constraints may also restrict the \emph{presentation surface} of returned data.
For example, a policy can disable access to sensitive fields; servers SHOULD implement deterministic redaction so that
clients can treat outputs as stable and safe to display.
When redaction occurs, servers SHOULD disclose which fields were redacted using non-sensitive identifiers (for example, field
paths) and SHOULD record redaction outcomes in audit events without logging raw sensitive values.
This behavior can be formalized as a function that returns both a presented object and a set of redacted field paths:
\begin{equation}
\mathrm{Present}(o, p) \rightarrow (o', F),
\label{eq:present}
\end{equation}
where $o$ is the raw domain object, $p$ is the effective policy, $o'$ is the redacted presentation, and $F$ is the set
of redacted fields.

\subsection{Tenant and Workspace Boundary Enforcement}
Tenant isolation is a protocol requirement, not an implementation detail.
In multi-tenant deployments, an integration app may be scoped to a specific workspace/organization.
Servers MUST enforce this boundary using server-side resolution (for example, by fetching resource metadata through adapters)
and MUST NOT trust client-provided tenant identifiers as the source of truth.
When policies restrict allowed organizations or resource IDs, those checks MUST compose with workspace boundaries.

\subsection{Reason Codes and Error Semantics}
OpenPort requires stable reason codes for denials so that agent runtimes can implement safe behavior (refresh discovery,
request approval, back off, or escalate to an operator).
Reason codes SHOULD be scoped and human-interpretable (for example, \texttt{agent.token\_invalid}, \texttt{agent.scope\_denied},
\texttt{agent.policy\_denied}, \texttt{agent.rate\_limited}), and MUST avoid embedding secrets or internal identifiers.
OpenPort standardizes a minimum set of \texttt{agent.*} codes for interoperability; implementations MAY extend with
additional codes as long as existing semantics remain stable.
For requests that fail before an integration app can be resolved (for example, missing/invalid token), implementations MAY
prefer aggregate security metrics over per-request audit events to avoid log amplification; for authenticated requests,
auditability is mandatory.

\subsection{Lifecycle: Issuance, Rotation, Revocation}
Production authorization requires full lifecycle support: key issuance, parallel keys for rotation, and revocation that
is effective immediately.
Credentials SHOULD support metadata useful for operations (token prefix, last-used timestamps, expiration) without
exposing secrets.

\subsubsection{Issuance}
An integration app is provisioned with an explicit scope set and policy constraints.
Keys are issued to apps and SHOULD be displayed only once at issuance; servers SHOULD store only a non-reversible form
of the secret (for example, a hash).

\subsubsection{Rotation}
Rotation SHOULD allow parallel validity windows (old and new keys) to avoid downtime and to support staged rollout.
Servers SHOULD provide stable key identifiers (not secrets) so operators can audit which credentials were used.

\subsubsection{Revocation and Emergency Disablement}
Revocation MUST take effect immediately across all endpoints, including discovery.
OpenPort treats revocation as a first-class security control: denials due to revocation MUST generate audit events with
stable reason codes.
Implementations SHOULD support an emergency disable switch at the app level to stop all agent access quickly during an
incident response.

\section{Authorization Flows}
This section specifies the expected end-to-end authorization flows that make OpenPort deployable in production.
The goal is not to prescribe an identity system, but to define an authorization \emph{closed loop}: issuance, use,
decision logging, and revocation.

\subsection{Provisioning, Issuance, and Tool Discovery (Integration Tokens)}
An operator provisions an integration app with explicit scopes and policy constraints, then issues an agent key.
The agent runtime uses the key to discover permitted tools via the manifest.
Discovery is part of the authorization surface: the manifest MUST be authorization-dependent (Eq.~\eqref{eq:visible-tools})
and MUST not leak tools or sensitive fields that are not permitted for the current integration.

\begin{figure}[t]
\centering
\begin{floatbox}
\begin{lstlisting}[basicstyle=\ttfamily\footnotesize]
Admin  -> Admin API: POST /api/agent-admin/v1/apps (scopes, policy)
Admin  -> Admin API: POST /api/agent-admin/v1/apps/{id}/keys (rotate/issue)
Agent  -> Agent API : GET /api/agent/v1/manifest (Bearer key)
Server -> Audit     : allow/deny + code + keyId/appId
\end{lstlisting}
\end{floatbox}
\caption{Integration token provisioning and authorized tool discovery.}
\label{fig:auth-discovery}
\end{figure}

\subsection{Authorized Reads and Policy Effects}
After discovery, the agent runtime calls read endpoints to retrieve data.
Read access is governed by the same decision algorithm (Eq.~\eqref{eq:allow}) and MUST enforce tenant/workspace
boundaries and ABAC constraints (resource allowlists, bounded query windows, and redaction).
When policies disable sensitive fields, servers SHOULD redact deterministically and SHOULD disclose redacted field paths
as in Eq.~\eqref{eq:present}.

\begin{figure}[t]
\centering
\begin{floatbox}
\begin{lstlisting}[basicstyle=\ttfamily\footnotesize]
Agent  -> Agent API : GET /api/agent/v1/ledgers
Server -> Audit     : allow/deny + code + resultCount
Agent  -> Agent API : GET /api/agent/v1/transactions?ledgerId=...
Server -> Audit     : allow/deny + code + redactedFields
\end{lstlisting}
\end{floatbox}
\caption{Authorized reads with policy enforcement and audit emission.}
\label{fig:read-flow}
\end{figure}

\subsection{Request Evaluation and Reason Codes}
For each request, OpenPort requires server-side evaluation in a fixed order: authentication (token validity), scope
matching, policy constraints (ABAC), and tenant/workspace boundary enforcement.
Each denial SHOULD map to a stable reason code (for example, \texttt{agent.token\_invalid}, \texttt{agent.token\_expired},
\texttt{agent.scope\_denied}, \texttt{agent.policy\_denied}, \texttt{agent.forbidden}, \texttt{agent.rate\_limited}).
These codes enable safe agent retry behavior, operator debugging, and security analytics without exposing sensitive
internal details.

\subsection{Write Requests: Draft-First and Optional Auto-Execute}
Write requests are submitted via \texttt{POST /api/agent/v1/actions}. OpenPort requires draft-first behavior: in the absence
of explicit, time-bounded policy enablement, write requests MUST produce drafts rather than execute side effects.
Auto-execute eligibility is governed by Eq.~\eqref{eq:autoexec} and is designed to be auditable and safely reversible
through revocation.

For high-risk tools, the agent runtime SHOULD call \texttt{POST /api/agent/v1/preflight} to obtain an impact summary and hash
(Eq.~\eqref{eq:preflight-hash}) before requesting execution.
If auto-execution is denied (disabled, expired, allowlist mismatch, missing justification, missing idempotency, or
preflight mismatch), the server MUST return a draft and a machine-readable denial code so operators can decide whether
to approve.
If an idempotency key is present and a prior execution exists for the same $(\mcode{appId}, \mcode{idempotencyKey})$,
the server SHOULD return the prior execution outcome (Eq.~\eqref{eq:idempotency-map}) to make retries safe.

\begin{figure}[t]
\centering
\begin{floatbox}
\begin{lstlisting}[basicstyle=\ttfamily\footnotesize]
Agent  -> Agent API : POST /api/agent/v1/preflight (action, payload)
Server -> Agent     : impact, impactHash, preflightId, stateWitnessHash?
Agent  -> Agent API : POST /api/agent/v1/actions (execute?, idem?, preflightHash?, preflightId?, stateWitnessHash?)
Server -> Agent     : Draft (default) or Execution (if allowed)
Server -> Audit     : allow/deny + code + draftId/executionId
\end{lstlisting}
\end{floatbox}
\caption{Write request flow with optional preflight and draft-first defaults.}
\label{fig:write-flow}
\end{figure}

\subsection{Admin Review and Execution}
Drafts provide a reviewable representation of write intent. The admin control plane lists drafts, inspects context
(including policy snapshots), and approves or rejects requests.
Approval triggers execution under the recorded policy constraints; rejection cancels the draft.
This human-in-the-loop step is a governance requirement for high-risk operations and is the primary mechanism to
mitigate prompt-injection-driven misuse.

OpenPort supports asynchronous client behavior by allowing agents to poll draft status:
\begin{lstlisting}[language=bash,caption={Draft review and polling endpoints (conceptual).}]
GET  /api/agent/v1/drafts/{id}
GET  /api/agent-admin/v1/drafts?status=draft
POST /api/agent-admin/v1/drafts/{id}/approve
POST /api/agent-admin/v1/drafts/{id}/reject
\end{lstlisting}
Clients SHOULD implement backoff for polling and treat drafts as non-final until an execution outcome is recorded.

\subsection{Rotation, Revocation, and Incident Response}
Authorization safety requires that credentials can be rotated and revoked without downtime.
Rotation issues a new key while keeping the old key valid for a controlled overlap window; revocation invalidates a key
or an entire app immediately across all endpoints, including \texttt{/manifest}.
Implementations SHOULD support emergency disablement at the app level to stop all agent traffic quickly during an
incident.

\begin{figure}[t]
\centering
\begin{floatbox}
\begin{lstlisting}[basicstyle=\ttfamily\footnotesize]
Admin -> Admin API: POST /api/agent-admin/v1/apps/{id}/keys (issue new key)
Agent -> Agent API : use new key (progressive rollout)
Admin -> Admin API: POST /api/agent-admin/v1/keys/{id}/revoke (revoke old key)
Agent -> Agent API : old key fails immediately (agent.token_invalid)
\end{lstlisting}
\end{floatbox}
\caption{Key rotation and immediate revocation.}
\label{fig:rotation-revocation}
\end{figure}

\subsection{Client Recovery Guidance}
Stable reason codes make agent runtimes safer by enabling deterministic recovery logic.
Table~\ref{tab:recovery} summarizes recommended responses for common denial codes.
These recommendations are intentionally aligned with the conformance invariants in Table~\ref{tab:invariants}: a server
that satisfies the invariants should enable clients to implement Table~\ref{tab:recovery} without relying on
model-specific heuristics.

\begin{table}[t]
\caption{OpenPort reason codes and recommended client behavior (baseline set with optional profile extensions).}
\label{tab:recovery}
\centering
\small
% Constrain width for single-column readability.
\begin{floatbox}
{\setlength{\tabcolsep}{4pt}
\renewcommand{\arraystretch}{1.05}
\begin{tabular}{>{\raggedright\arraybackslash}p{0.34\linewidth} >{\raggedright\arraybackslash}p{0.60\linewidth}}
\toprule
\textbf{Reason code} & \textbf{Recommended client action} \\
\midrule
\url{agent.token_invalid}\newline\url{agent.token_expired} & Stop; request a new key/token; refresh discovery after re-authentication. \\
\url{agent.scope_denied} & Refresh \texttt{/manifest}; request additional scopes only if explicitly approved. \\
\url{agent.policy_denied}\newline\url{agent.forbidden} & Surface to operator; adjust query/window/resource selection; do not retry blindly. \\
\url{agent.action_unknown}\newline\url{agent.action_invalid} & Do not retry blindly; refresh \texttt{/manifest} and validate tool name and input schema. \\
\url{agent.preflight_required}\newline\url{agent.preflight_mismatch}\newline\url{agent.preflight_not_found} & Run \texttt{/preflight} and retry with returned hash (or a valid \texttt{preflightId}); treat mismatch as a change in impact and do not execute blindly. \\
\url{agent.precondition_failed}\newline\textit{(State Witness profile)} & Treat as state change (TOCTOU); rerun \texttt{/preflight} to obtain a fresh witness/hash and require renewed operator approval before execution. \\
\url{agent.idempotency_required} & Generate a stable idempotency key per write intent and retry once. \\
\url{agent.idempotency_replay} & Treat as success; reuse returned execution outcome; do not re-execute. \\
\url{agent.auto_execute_disabled}\newline\url{agent.auto_execute_expired}\newline\url{agent.auto_execute_denied} & Accept draft result; route to admin approval rather than retrying execution. \\
\url{agent.draft_not_found}\newline\url{agent.draft_already_final} & Stop; refresh draft status from the gateway; treat as non-retriable and require operator review. \\
\url{agent.step_up_required}\newline\url{agent.step_up_invalid} & Complete step-up verification through the operator channel and retry within the allowed window. \\
\url{agent.rate_limited} & Back off; respect \texttt{Retry-After} if present; reduce polling; consider caching \texttt{/manifest}. \\
\bottomrule
\end{tabular}
}
\end{floatbox}
\end{table}

\subsection{Delegated Authorization Binding (OAuth 2.0)}
OpenPort is compatible with delegated ``act on behalf of'' use cases, commonly implemented via OAuth 2.0~\cite{rfc6749}.
However, delegated identity is insufficient without governance:
\begin{itemize}
  \item delegated tokens MUST map to minimal OpenPort scopes;
  \item delegated sessions MUST apply policy windows (time bounds, resource sets, field redaction);
  \item revocation MUST be effective immediately (including user consent withdrawal).
\end{itemize}
The binding MAY use token introspection or local validation, but the resulting authorization decision MUST be enforced
server-side and MUST be auditable.

\begin{figure}[t]
\centering
\begin{floatbox}
\begin{lstlisting}[basicstyle=\ttfamily\footnotesize]
User  -> IdP       : OAuth authorize (consent)
Agent -> IdP       : Obtain delegated access token
Agent -> Agent API : Call tool (delegated token)
Server-> AuthZ     : Map claims -> scopes + policy window
Server-> Audit     : allow/deny + reason + subject + appId
\end{lstlisting}
\end{floatbox}
\caption{Delegated authorization binding: identity plus governance.}
\label{fig:delegated-binding}
\end{figure}

\section{Risk-Gated Writes}
Write operations are the primary risk surface for agent tooling because they create side effects and are harder to
contain than reads.
OpenPort therefore treats ``write access'' as a governed capability rather than a default API behavior.
The protocol's baseline write posture is \textbf{draft-first}: write requests create reviewable drafts unless
auto-execution is explicitly enabled and time-bounded.

\subsection{Draft-First Semantics}
Write requests are submitted via \texttt{POST /api/agent/v1/actions} using a tool name and a payload (schemas are
discovered via the manifest).
The server MUST evaluate authorization and policy constraints server-side (Eq.~\eqref{eq:allow}) before any side effects
can occur.
Authorization failures (invalid token, missing scope, policy denial) remain hard errors with stable reason codes.
If the request is authorized but not eligible for immediate execution, the server MUST return a draft.
If execution was explicitly requested (for example, \texttt{execute=true}) but denied by auto-execute policy, the server MUST
attach a stable machine-readable denial reason code for the \emph{auto-execute request}.
Clients MAY explicitly force drafting (for example, \texttt{forceDraft=true}) to require operator review even when an
auto-execute window is enabled.
This ``fail closed into draft'' rule converts many unsafe behaviors (tool misuse, prompt injection, retries) into a
review queue rather than side effects.

\begin{figure}[t]
\centering
\begin{floatbox}
\begin{lstlisting}[basicstyle=\ttfamily\footnotesize]
Agent  -> Agent API: POST /api/agent/v1/actions (tool, args, execute?)
Server -> Audit    : allow/deny + code + risk tier + draftId
Server -> Agent    : Draft (default) or Execution (if allowed)
Admin  -> Admin API: POST /api/agent-admin/v1/drafts/{id}/approve
Server -> Target   : Execute (idempotency + optional preflight hash)
Server -> Audit    : execution success/fail + executionId
\end{lstlisting}
\end{floatbox}
\caption{Draft-first write pipeline with approval and high-risk safeguards.}
\label{fig:draft-pipeline}
\end{figure}

\subsection{Draft and Execution Objects}
OpenPort separates \emph{intent} (draft) from \emph{effect} (execution). A draft represents a write request that can be
reviewed and approved by an operator. An execution represents the outcome of running a tool under enforced constraints.

At minimum, a draft record SHOULD include:
\begin{itemize}
  \item identifiers: \texttt{draftId}, \texttt{appId}, \texttt{keyId}, actor identity;
  \item intent: tool name (\texttt{actionType}) and payload;
  \item governance: risk tier, \texttt{auto\_execute\_requested}, optional \texttt{justification};
  \item safeguards: optional \texttt{preflight} impact summary and \texttt{preflightHash}, optional \texttt{idempotencyKey};
  \item policy snapshot inputs (for example, required scopes and auto-execute configuration) to support review and incident
  response.
\end{itemize}
Executions SHOULD include an \texttt{executionId}, the associated \texttt{draftId}, a terminal status, and either a
result object or an error string, plus timestamps for audit correlation.

\subsection{Draft Lifecycle and State Machine}
Drafts are long-lived governance objects and therefore require explicit lifecycle states.
The reference runtime uses a small state machine aligned with operational workflows.
Table~\ref{tab:draft-lifecycle} summarizes the semantics and transitions.

\begin{table}[t]
\caption{Draft lifecycle states and transitions (reference profile).}
\label{tab:draft-lifecycle}
\centering
\small
% Constrain width for single-column readability.
\begin{floatbox}
{\setlength{\tabcolsep}{4pt}
\renewcommand{\arraystretch}{1.05}
\begin{tabular}{>{\raggedright\arraybackslash}p{0.18\linewidth} >{\raggedright\arraybackslash}p{0.42\linewidth} >{\raggedright\arraybackslash}p{0.32\linewidth}}
\toprule
\textbf{Status} & \textbf{Meaning} & \textbf{Transitions (trigger)} \\
\midrule
\texttt{draft} & Pending operator decision; no side effects have been executed. & \texttt{confirmed} (approve or eligible auto-execute), \texttt{canceled} (reject). \\
\texttt{confirmed} & Approved for execution; a successful execution outcome should be recorded for completion. & \texttt{failed} (execution attempt fails). \\
\texttt{canceled} & Rejected by operator; terminal. & none. \\
\texttt{failed} & Execution attempted and failed; terminal. & none. \\
\bottomrule
\end{tabular}
}
\end{floatbox}
\end{table}

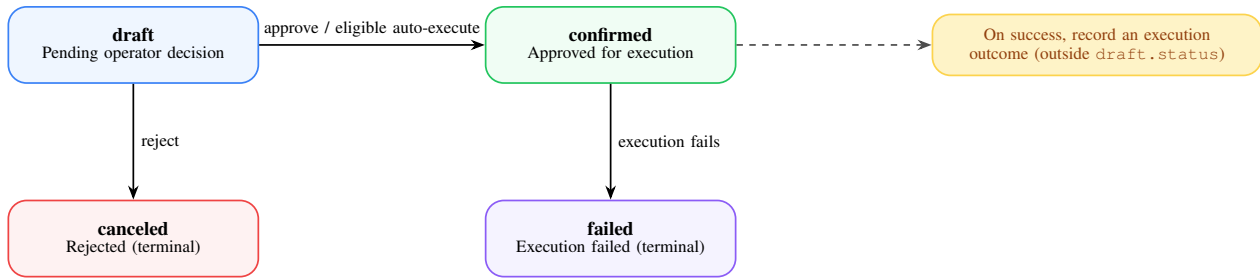
\begin{figure}[t]
\centering
\begin{floatbox}
\resizebox{\linewidth}{!}{%
\begin{tikzpicture}[
  >=Stealth,
  state/.style={draw, rounded corners=3mm, thick, align=center, minimum width=4.1cm, minimum height=1.25cm, inner sep=5pt, font=\small},
  note/.style={draw=opAmber, fill=opAmberFill, text=opAmberText, rounded corners=3mm, thick, align=center, minimum width=5.4cm, minimum height=1.0cm, inner sep=5pt, font=\small},
  edge/.style={->, thick},
  edgedash/.style={->, thick, dashed, color=black!70},
  elabel/.style={font=\footnotesize, align=center}
]
\node[state, draw=opBlue, fill=opBlueFill] (draft) {\textbf{draft}\\[-2pt]\footnotesize Pending operator decision};
\node[state, draw=opGreen, fill=opGreenFill, right=3.7cm of draft] (confirmed) {\textbf{confirmed}\\[-2pt]\footnotesize Approved for execution};
\node[state, draw=opRed, fill=opRedFill, below=1.9cm of draft] (canceled) {\textbf{canceled}\\[-2pt]\footnotesize Rejected (terminal)};
\node[state, draw=opPurple, fill=opPurpleFill, below=1.9cm of confirmed] (failed) {\textbf{failed}\\[-2pt]\footnotesize Execution failed (terminal)};
\node[note, right=3.2cm of confirmed] (note) {\footnotesize On success, record an execution\\[-1pt]\footnotesize outcome (outside \texttt{draft.status})};

\draw[edge] (draft) -- node[elabel, above] {approve / eligible auto-execute} (confirmed);
\draw[edge] (draft) -- node[elabel, right] {reject} (canceled);
\draw[edge] (confirmed) -- node[elabel, right] {execution fails} (failed);
\draw[edgedash] (confirmed) -- (note);
\end{tikzpicture}%
}
\end{floatbox}
\caption{Draft lifecycle state machine (reference profile).}
\label{fig:draft-lifecycle}
\end{figure}

Figure~\ref{fig:draft-lifecycle} visualizes the same draft state machine.

Clients SHOULD treat \texttt{draft.status} as governance state, not as a substitute for an execution record.
For asynchronous operation, clients MAY poll \texttt{GET /api/agent/v1/drafts/\{id\}} to retrieve the latest draft state
and the most recent execution record.

\subsection{High-Risk Safeguards: Preflight Hashing and Idempotency}
For high-risk tools, OpenPort supports additional safeguards that make writes predictable under retries and reduce the
chance of unintended side effects.
These safeguards are used as gates for auto-execution and as evidence for operator review.

\subsubsection{Preflight Impact Hash}
The \texttt{POST /api/agent/v1/preflight} endpoint computes an impact summary and returns an impact hash.
The preflight hash binds the action, payload, and impact summary using Eq.~\eqref{eq:preflight-hash}.
Clients supply this value as \texttt{preflightHash} in subsequent \texttt{POST /api/agent/v1/actions} requests.
When preflight is required for auto-execution, the server MUST deny auto-execution unless the client supplies a matching
hash; denial codes include \url{agent.preflight_required} and \url{agent.preflight_mismatch}. When a deployment offers a
preflight handle (\texttt{preflightId}) and the handle cannot be resolved, it SHOULD fail closed with
\url{agent.preflight_not_found} rather than executing with a regenerated payload.
This makes high-risk writes robust to agent retries and ``time-of-check/time-of-use'' confusion: the execution request
is tied to the reviewed impact summary, not to an implicit UI interpretation.

\subsubsection{Idempotency and Replay}
Idempotency converts an unsafe retry pattern into a deterministic map (Eq.~\eqref{eq:idempotency-map}).
If a request carries an idempotency key and an execution already exists for the same
$(\mcode{appId}, \mcode{idempotencyKey})$, the server SHOULD return the prior execution outcome and MUST NOT re-execute
side effects.
The reference profile uses \url{agent.idempotency_replay} for this case, enabling clients to treat the response as a
success path.
If high-risk auto-execution is enabled but an idempotency key is missing, the server MUST fail closed into a draft and
return \url{agent.idempotency_required}.

\subsection{Time-Bounded Auto-Execute (Optional)}
Auto-execution is an explicit capability that MAY be enabled for a limited time window and tool allowlist.
Eligibility is defined by Eq.~\eqref{eq:autoexec}.
If an agent requests execution but $\mathrm{AutoExecAllowed}(r)$ is false, the server MUST return a draft and MUST return
a stable denial reason code.
In the reference implementation, common denial codes include \url{agent.auto_execute_disabled},
\url{agent.auto_execute_expired}, and \url{agent.auto_execute_denied}; high-risk guard failures use
\url{agent.preflight_required}, \url{agent.preflight_mismatch}, and \url{agent.idempotency_required}.
For high-risk auto-execution, the profile also requires a non-empty justification; missing justification is treated as
invalid input and returns \url{agent.action_invalid}.

\begin{lstlisting}[basicstyle=\ttfamily\footnotesize,caption={Server-side decision procedure for \texttt{POST /actions} (conceptual).}]
if execute && !forceDraft:
  if idempotencyKey && E(app,idempotencyKey) exists:
    return executed(replayed=true)
compute impact (+ preflight hash) for high-risk tools
denial = AutoExecAllowed(request) ? null : <reason code>
draft = saveDraft(status = denial ? "draft" : "confirmed", snapshot=policyInputs)
audit(draft, denial)
if denial: return draft(denial)
execution = executeDraft(draft)
return executed(execution)
\end{lstlisting}

\subsection{Operator Approval and Separation of Duties}
Draft approval and rejection are performed through an admin control plane (for example,
\texttt{POST /api/agent-admin/v1/drafts/\{id\}/approve} and \texttt{/reject}).
This separation of duties is intentional: the same credentials used for agent access MUST NOT be sufficient to approve
high-risk writes.
Approval triggers execution with full audit correlation (draftId, executionId, operator identity), and rejection
transitions the draft to \texttt{canceled}.
Attempts to act on missing or finalized drafts SHOULD return stable denial codes such as
\url{agent.draft_not_found} and \url{agent.draft_already_final}.

\section{Rate Limits and Abuse Controls}
Rate limiting is the operational ``safety valve'' for exposed agent tooling.
Even with correct scopes and policies, a compromised key or a runaway agent loop can amplify cost and availability
impact through excessive calls, aggressive polling, or repeated retries.
OpenPort therefore treats admission control as part of the protocol's governance surface.

\subsection{Objectives}
OpenPort rate limiting and abuse controls target three production objectives:
\begin{itemize}
  \item \textbf{Bound blast radius under leakage}: constrain the maximum request volume per credential and per network
  origin, so a stolen key cannot rapidly enumerate or exfiltrate data.
  \item \textbf{Prevent cost amplification}: protect expensive endpoints (for example, \texttt{/actions} and \texttt{/preflight})
  from retry storms and polling loops.
  \item \textbf{Keep behavior predictable}: return stable, machine-actionable semantics so agent runtimes can implement
  deterministic recovery (Table~\ref{tab:recovery}) and conformance tests can verify consistent 429 behavior
  (Table~\ref{tab:invariants}).
\end{itemize}

\subsection{Rate Limit Dimensions and Default Policy}
At minimum, OpenPort implementations MUST enforce rate limits by the \emph{(credential, client IP)} tuple.
The credential dimension prevents a single integration key from overwhelming the system; the IP dimension reduces abuse
from shared or spoofed credentials and provides a coarse network control.
More granular quotas are OPTIONAL but recommended for real deployments:
\begin{itemize}
  \item per integration app (\texttt{appId}) to support multiple keys and rotations;
  \item per tenant/workspace/resource to prevent cross-tenant ``fan out'' even under valid scopes;
  \item per tool/action to assign lower budgets to high-cost operations;
  \item per endpoint class (discovery, reads, writes, polling) so that \texttt{/manifest} and \texttt{/drafts} do not
  consume the same budget as \texttt{/transactions}.
\end{itemize}
OpenPort does not mandate a specific algorithm, but it requires predictable 429 behavior across these dimensions.
Admission control MUST be evaluated \emph{before} executing any tool logic and before allocating durable governance
objects (drafts, executions). In particular, rate-limited requests MUST NOT create drafts or side effects.
More generally, let $\mathcal{G}$ denote the set of governance objects created by the write pipeline (drafts and
executions). For a request $r$, define $\Delta \mathcal{G}(r)$ as the newly allocated governance objects during handling
of $r$. OpenPort's hard-denial posture is:
\begin{equation}
\Delta \mathcal{G}(r)=\emptyset \quad \text{whenever } \neg \mathrm{Allow}(r),
\label{eq:no-alloc-on-deny}
\end{equation}
which formalizes ``no drafts on denials'': authentication failures, scope/policy denials, and admission-control denials
must not allocate drafts or executions.

\subsection{Fixed-Window Admission Rule}
One simple and testable limiter is a fixed-window admission rule.
For a given key identifier $k$ and client IP $u$, let $N_{k,u}(t)$ be the number of accepted requests in the interval
$[t-W, t)$ where $W$ is the window size.
The limiter admits a request at time $t$ iff:
\begin{equation}
N_{k,u}(t) < L,
\label{eq:fixed-window}
\end{equation}
where $L$ is the maximum number of requests per window.
Token-bucket-style mechanisms~\cite{rfc2697}, sliding-window, and multi-dimensional quota systems are compatible, provided
they preserve stable 429 semantics and can be audited.

\subsection{429 Semantics and Client Backoff}
When a request is denied due to admission control, the server SHOULD return HTTP 429 with a stable code
\url{agent.rate_limited} and a parseable error envelope.
If available, servers SHOULD include \texttt{Retry-After} guidance~\cite{rfc9110,rfc6585}; otherwise, clients MUST apply
a conservative backoff strategy (for example, exponential backoff with jitter) rather than retrying immediately.
Clients SHOULD also:
\begin{itemize}
  \item reduce polling frequency for draft status and admin queues;
  \item cache \texttt{/manifest} and avoid re-discovery on every call;
  \item reuse \texttt{idempotencyKey} on write retries so a delayed retry cannot duplicate side effects
  (Eq.~\eqref{eq:idempotency-map}).
\end{itemize}
Stable rate-limit behavior is a governance requirement because it determines whether agent runtimes converge to safe
behavior under pressure, or amplify load by retrying unpredictably.

\begin{lstlisting}[language=json,caption={Example 429 response envelope.}]
{ "ok": false, "code": "agent.rate_limited", "message": "Rate limit exceeded" }
\end{lstlisting}

\subsection{Additional Abuse Controls}
Rate limits are necessary but insufficient. OpenPort pairs admission control with additional ``abuse hardening''
requirements:
\begin{itemize}
  \item \textbf{Network policy allowlists}: optional IP allowlists deny traffic outside expected network boundaries with a
  policy denial code.
  \item \textbf{Schema validation and bounded parsing}: invalid inputs MUST fail as 4xx errors with stable codes (for example,
  unknown action or invalid payload), and SHOULD NOT degrade into 5xx failures.
  \item \textbf{Emergency disablement}: operators SHOULD be able to disable an integration app quickly during incident
  response, cutting off traffic across all endpoints including discovery.
\end{itemize}
These controls are designed to be testable in black-box conformance harnesses and to integrate cleanly with audit trails.

\subsection{Reference Profile Parameters}
The reference runtime enforces rate limiting at the authentication layer, before executing any endpoint logic.
This ensures uniform protection across \texttt{/api/agent/v1/*} endpoints, including discovery and preflight.
The current reference profile applies Eq.~\eqref{eq:fixed-window} per \{keyId, client IP\} with $W=60$ seconds and
$L=240$ requests per window, using a bucket key of the form \texttt{agent:\{keyId\}:\{ip\}}.

\section{Auditability}
Auditability is mandatory for safe agent tool exposure. Without an audit trail, operators cannot answer the core
production questions: \emph{what} was accessed, \emph{what} was attempted, \emph{which} credential was used, \emph{which}
policy decision was applied, and \emph{what} side effects occurred.
OpenPort therefore treats audit events as first-class protocol outputs: they are required for incident response,
compliance, and reproducible debugging of agent behavior.

\subsection{Goals}
OpenPort audit requirements target three concrete goals:
\begin{itemize}
  \item \textbf{Accountability}: attribute requests and side effects to an integration app/key, an actor identity, and
  (when applicable) an approving operator.
  \item \textbf{Reconstructability}: enable an operator to reconstruct request intent, authorization decision, and write
  lifecycle from drafts to executions, including idempotent replays and preflight binding.
  \item \textbf{Operability}: provide stable machine-facing reason codes for denial paths so incidents can be triaged and
  automated recovery strategies can be implemented safely (Table~\ref{tab:recovery}).
\end{itemize}

\subsection{Event Taxonomy}
OpenPort requires audit events for \textbf{allow}, \textbf{deny}, and \textbf{fail} paths.
The protocol does not mandate an internal logging system, but it does require a stable event taxonomy that is
independent of model runtime behavior.
The reference profile uses dot-separated action identifiers to group events:
\begin{itemize}
  \item \textbf{Agent data reads}: for example, \url{agent.ledger.list}, \url{agent.transaction.list}.
  \item \textbf{Agent write governance}: for example, \url{agent.action.preflight}, \url{agent.action.draft.created},\newline
  \url{agent.action.auto_execute.requested}, \url{agent.action.execute}, \url{agent.action.idempotency_replay}.
  \item \textbf{Admin control plane}: for example, \url{agent_app.create}, \url{agent_key.create},
  \url{agent_app.policy.update}, \url{agent_app.auto_execute.update},\newline \url{agent_key.revoke},
  \url{agent_app.revoke}, \url{agent.draft.approve}, \url{agent.draft.reject}.
  \item \textbf{Protective controls}: rate-limit denials and policy denials, recorded with stable reason codes such as
  \url{agent.rate_limited}, \url{agent.token_invalid}, \url{agent.token_expired}, \url{agent.scope_denied}, and
  \url{agent.policy_denied}.
\end{itemize}
Implementations MAY extend the action namespace, but they SHOULD preserve backward compatibility for core actions to
support detection rules and long-lived operational dashboards.

\subsection{Minimum Event Schema and Correlation}
Audit events MUST be structured and machine-parseable. The reference runtime uses an event schema with the following
minimum fields:
\begin{table}[t]
\caption{Minimum audit event fields (reference profile).}
\label{tab:audit-fields}
\centering
\small
% Constrain width for single-column readability.
\begin{floatbox}
{\setlength{\tabcolsep}{4pt}
\renewcommand{\arraystretch}{1.05}
\begin{tabular}{>{\raggedright\arraybackslash}p{0.28\linewidth} >{\raggedright\arraybackslash}p{0.66\linewidth}}
\toprule
\textbf{Field} & \textbf{Meaning / usage} \\
\midrule
\texttt{id}, \texttt{created\_at} & Server-generated event identifier and timestamp for ordering and retention. \\
\texttt{action}, \texttt{status} & Stable event name and outcome (\texttt{success}, \texttt{denied}, \texttt{failed}). \\
\texttt{code} & Stable reason code for denials (for example, \url{agent.scope_denied}) and for protection outcomes (for example, \url{agent.rate_limited}). \\
\texttt{app\_id}, \texttt{key\_id} & Integration identifiers for attribution, revocation correlation, and blast-radius analysis. \\
\texttt{actor\_user\_id} & Subject identity under which domain reads/writes are performed (service user or user). \\
\texttt{performed\_by\_user\_id} & Approving operator identity for admin actions and write approvals (separation of duties). \\
\texttt{request\_id} & Optional client correlation identifier to tie retries and multi-step sequences. \\
\texttt{draft\_id}, \texttt{execution\_id} & Write lifecycle correlation from intent (draft) to effect (execution). \\
\texttt{ip}, \texttt{user\_agent} & Request metadata for abuse analytics, incident response, and allowlist debugging. \\
\texttt{details} & Bounded, redacted metadata (risk tier, tool name, preflight summary, redacted fields), excluding secrets. \\
\bottomrule
\end{tabular}
}
\end{floatbox}
\end{table}

Correlation is a protocol design requirement: drafts and executions MUST be linkable to the audit trail.
Fields MAY be null when they are not available (for example, invalid or missing credentials may prevent resolving
\texttt{app\_id} and \texttt{key\_id}). In these cases, events SHOULD still capture request metadata and stable reason
codes to support abuse detection and incident response.
In particular:
\begin{itemize}
  \item \texttt{requestId} (if supplied) SHOULD propagate into audit to correlate agent retries and preflight/action pairs.
  \item \texttt{draftId} MUST be recorded for draft creation, approval/rejection, and execution outcomes.
  \item \texttt{executionId} MUST be recorded for execution outcomes and for idempotent replays.
\end{itemize}
To make write outcomes reconstructable from audit alone, OpenPort also treats draft-to-execution linking as a protocol
invariant. Let $\mathcal{D}$ be the set of drafts and $\mathcal{E}$ be the set of execution records. OpenPort requires:
\begin{equation}
\forall e \in \mathcal{E}\colon \exists! d \in \mathcal{D}\ \text{s.t.}\ \mathsf{draftId}(e)=\mathsf{id}(d),
\label{eq:draft-exec-link}
\end{equation}
where $\exists!$ denotes existence and uniqueness, $\mathsf{draftId}(e)$ denotes the draft identifier carried by execution
record $e$, and $\mathsf{id}(d)$ denotes the server-issued identifier of draft record $d$.
In words: every execution record MUST reference exactly one recorded draft intent.
This invariant ensures every execution can be traced back to a single policy snapshot and approval decision, and it gives
clients and auditors a deterministic join key between intent and effects.

Operationally, Eq.~\eqref{eq:draft-exec-link} is a referential-integrity requirement on the write pipeline.
The extraction function $\mathsf{draftId}(\cdot)$ reads the draft identifier field from an execution record (as emitted by
the gateway), and $\mathsf{id}(\cdot)$ is the primary identifier of a persisted draft record.
The uniqueness clause rules out ambiguous correlation under retries or duplicate drafts, and the existence clause rules
out ``side-effect'' executions that bypass draft creation and approval entirely.

\begin{lstlisting}[language=json,caption={Example audit event (conceptual).}]
{
  "id": "aud_...",
  "created_at": "2026-02-15T00:00:00Z",
  "action": "agent.action.draft.created",
  "status": "denied",
  "code": "agent.auto_execute_disabled",
  "app_id": "app_...",
  "key_id": "key_...",
  "request_id": "req_...",
  "draft_id": "drf_...",
  "ip": "203.0.113.4",
  "details": { "actionType": "transaction.hard_delete", "risk": "high" }
}
\end{lstlisting}

\subsection{Data Minimization and Redaction}
Audit events are security artifacts and MUST minimize sensitive content. OpenPort requires:
\begin{itemize}
  \item \textbf{No secrets in audit}: bearer tokens, refresh tokens, and raw credentials MUST NOT appear in \texttt{details}
  or free-form error strings.
  \item \textbf{Bounded details}: \texttt{details} SHOULD be size-bounded and SHOULD avoid full request payloads. For
  example, transaction exports should audit row counts and policy parameters, not raw exported data.
  \item \textbf{Redaction awareness}: when data policies redact fields (Eq.~\eqref{eq:present}), audit events SHOULD record
  redacted field paths and policy mode, without recording raw sensitive values.
  \item \textbf{Sanitized failures}: adapter- or domain-level execution failures SHOULD be mapped to safe error classes;
  raw stack traces and high-entropy payload fragments SHOULD NOT be recorded in audit sinks.
\end{itemize}
Operational logs MAY contain richer debugging context, but they SHOULD be separated from immutable audit sinks and
protected with different access controls.

\subsection{Integrity, Retention, and Access Control}
Audit sinks SHOULD be append-only and durable (for example, WORM storage or an external SIEM pipeline) to support incident
response and compliance.
Implementations SHOULD support integrity protections (for example, event signing or hash chaining) to make tampering detectable,
for example via transparency-log style, append-only Merkle tree constructions~\cite{rfc6962}.
Access to audit streams MUST be restricted to operators and automated security systems; agent credentials MUST NOT be
sufficient to read the audit stream.
The reference profile exposes audit listing only through the admin plane (for example, \texttt{GET /api/agent-admin/v1/audit})
for demonstration; production deployments SHOULD export audit events to an external sink and SHOULD avoid making the
entire audit stream queryable via a public-facing API.

\section{Normative Requirements and Verifiable Invariants}
\label{sec:invariants}
OpenPort expresses requirements using RFC 2119/8174 keywords (MUST/SHOULD/MAY)~\cite{rfc2119,rfc8174}.
In deployments, ``conformance'' is only meaningful if requirements are (i) verifiable from the outside and
(ii) stable over time.
OpenPort therefore defines a set of externally observable invariants and encourages implementations to publish which
conformance profile(s) they satisfy.

\subsection{Normative Language and Compliance Claims}
Key normative requirements include:
\begin{itemize}
  \item \textbf{MUST} enforce deny-by-default scope checks and server-side tenant boundaries.
  \item \textbf{MUST} support immediate revocation for keys and apps (including discovery).
  \item \textbf{MUST} default writes to drafts; auto-execute must be explicitly enabled and time-bounded.
  \item \textbf{MUST} return stable response envelopes and stable reason codes for denials and protective controls.
  \item \textbf{MUST} emit audit events for allow/deny/fail paths with stable reason codes.
  \item \textbf{MUST} implement operational rate limiting.
  \item \textbf{SHOULD} require preflight and idempotency for high-risk execution.
\end{itemize}
A deployment that claims conformance to an OpenPort profile MUST satisfy every requirement of that profile.
Profiles are versioned: tightening semantics or expanding required behavior SHOULD create a new profile identifier rather
than silently changing the meaning of an existing claim.
Backward compatibility is defined in operational terms: stable envelopes, stable reason-code semantics, and stable
discovery behavior under unchanged authorization inputs.

\subsection{Conformance Profiles}
OpenPort profiles are machine-readable descriptions of the protocol surface a server exposes.
A minimal profile typically specifies:
(i) required endpoints (method + path),
(ii) response envelope requirements for success and error cases, and
(iii) security minimums (for example, unauthenticated discovery is denied).
The reference repository provides a minimal core profile and an executable runner that can validate a local reference
runtime or a remote deployment.

\begin{lstlisting}[language=json,caption={Conceptual conformance profile fragment.}]
{
  "requiredEndpoints": [
    { "method": "GET",  "path": "/api/agent/v1/manifest" },
    { "method": "POST", "path": "/api/agent/v1/actions" }
  ],
  "envelope": {
    "success": { "requiredFields": ["ok","code","data"] },
    "error":   { "requiredFields": ["ok","code","message"] }
	  }
}
\end{lstlisting}

Profiles SHOULD be conservative and domain-neutral: they should not assume any specific product schema, and they should
avoid irreversible operations.
Draft-first writes make conformance safe because write intents can be exercised without applying side effects.
For production deployments, OpenPort SHOULD be validated via layered profiles:
\textbf{core} (endpoint presence + envelope consistency),
\textbf{authz} (scope/policy denials + revocation behavior),
\textbf{writes} (draft-first and high-risk safeguards), and
\textbf{abuse} (rate limits and non-5xx fuzz regression).
The reference repository includes a minimal core runner; the remaining profiles are intended as extensions that gate
high-stakes deployments.

\subsection{Invariants as Executable Checks}
This paper uses ``invariant'' in a strict sense: a statement about observable behavior that can be checked from the
outside using only protocol interactions and stable reason codes.
Stable reason codes are the shared interface between verification and operations: the same codes that drive client
recovery (Table~\ref{tab:recovery}) enable deterministic test oracles and predictable incident response.

Several OpenPort properties are amenable to lightweight formalization. The equations introduced earlier are not merely
expository; they define executable properties:
\begin{itemize}
  \item authorization-dependent discovery (Eq.~\eqref{eq:visible-tools})
  \item execution binding via preflight hash (Eq.~\eqref{eq:preflight-hash})
  \item TOCTOU resistance via state witness preconditions (Eq.~\eqref{eq:state-witness-hash})
  \item replay safety via idempotency mapping (Eq.~\eqref{eq:idempotency-map})
  \item deterministic allow/deny semantics and reason-code stability (Eq.~\eqref{eq:allow}, Eq.~\eqref{eq:first-fail})
  \item bounded query windows and deterministic presentation (Eq.~\eqref{eq:max-days}, Eq.~\eqref{eq:present})
  \item rate limiting as a testable admission rule (Eq.~\eqref{eq:fixed-window})
  \item explicit time-bounded auto-execute eligibility (Eq.~\eqref{eq:autoexec})
\end{itemize}
Table~\ref{tab:invariants} summarizes representative invariants and black-box checks.

\begin{table}[t]
\caption{Representative invariants and black-box checks.}
\label{tab:invariants}
\centering
\small
% Constrain width for single-column readability.
\begin{floatbox}
{\setlength{\tabcolsep}{4pt}
\renewcommand{\arraystretch}{1.05}
\begin{tabular}{>{\raggedright\arraybackslash}p{0.34\linewidth} >{\raggedright\arraybackslash}p{0.60\linewidth}}
\toprule
\textbf{Invariant (reference)} & \textbf{Black-box check (observable)} \\
\midrule
Envelope stability &
All endpoints return the standard envelope; errors are parseable and include a stable \texttt{code}. \\
AuthZ-dependent discovery (Eq.~\eqref{eq:visible-tools}) &
With insufficient scopes/policy, \texttt{/manifest} omits restricted tools/fields; adding scopes changes the visible set without breaking schema. \\
Immediate revocation &
After revocation, requests across endpoints (including \texttt{/manifest}) return 401 with a stable token denial code (for example, \url{agent.token_invalid}). \\
Allow/deny determinism (Eq.~\eqref{eq:allow}, Eq.~\eqref{eq:first-fail}) &
For a given request shape, denial code matches the first failing predicate in the prescribed order; codes remain stable across versions. \\
Preflight binding (Eq.~\eqref{eq:preflight-hash}) &
When preflight is required, missing hash yields a required denial; mismatched hash yields a mismatch denial; an unresolvable \texttt{preflightId} yields \url{agent.preflight_not_found}; matching hash permits progression to draft/execute. \\
State witness preconditions (Eq.~\eqref{eq:state-witness-hash}; optional profile) &
If a draft is bound to \texttt{stateWitnessHash}, execution revalidates current witness state; mismatch yields 409 with \url{agent.precondition_failed} and no side effects. \\
Idempotency (Eq.~\eqref{eq:idempotency-map}) &
Two identical write intents with the same \texttt{idempotencyKey} return the same execution record and do not produce duplicate side effects. \\
Auto-execute gating (Eq.~\eqref{eq:autoexec}) &
Execution is denied outside the configured window, without allowlist membership, or without required justification/guards; in these cases, a draft is returned with a denial code. \\
Rate limiting (Eq.~\eqref{eq:fixed-window}) &
Exceeding the configured limit yields 429 with a stable \url{agent.rate_limited} code; the request does not create drafts or side effects. \\
Audit completeness (Table~\ref{tab:audit-fields}) &
Allow/deny/fail paths emit structured audit events with correlation identifiers; audit payloads do not contain secrets. \\
\bottomrule
\end{tabular}
}
\end{floatbox}
\end{table}

These checks form a minimal ``safety bar'' for production exposure. Importantly, they are domain-neutral: they can be
run against any OpenPort deployment regardless of its underlying application schema, because they rely on the gateway's
governance semantics rather than product-specific logic.

\subsection{From Invariants to Release Gates}
In an open-source governance protocol, invariants act as regression budgets: changes are acceptable only if they
preserve conformance and do not weaken security properties.
The reference repository operationalizes this through a release-gate pipeline that combines build and unit tests,
conformance checks, fuzz/abuse regression (``no 5xx'' for malformed inputs), and safety scans to prevent accidental
secrets or boundary leakage in public artifacts.
This engineering discipline is part of the protocol's governance story: it makes the specification enforceable in
practice and reduces the chance that future releases silently expand the agent's authority surface.

\section{Reference Implementation and Engineering Methodology}
We provide a reference runtime to validate the specification and reduce adoption cost.
The implementation is intentionally small and deterministic: it exists to make OpenPort's governance semantics executable,
not to prescribe an application architecture.
This section describes the reference boundary (what is standardized vs what remains behind adapters) and the engineering
method used to ship the protocol safely as open source.

\subsection{Reference Runtime Goals}
The reference runtime is designed around four concrete goals:
\begin{itemize}
  \item \textbf{Make governance semantics concrete}: implement the agent-facing endpoints, stable envelopes, and reason
  codes; enforce scopes and ABAC constraints; and implement the draft-first write pipeline (preflight hashing,
  idempotency, and optional time-bounded auto-execute).
  \item \textbf{Remain domain-neutral}: keep product schemas and business rules behind adapters so that adopting OpenPort
  does not require open-sourcing internal models.
  \item \textbf{Fail closed}: deny-by-default on discovery and execution; ensure denials do not allocate durable governance objects
  (drafts/executions) and do not execute domain side effects; return stable \texttt{agent.*} reason codes for safe client
  behavior.
  \item \textbf{Be externally verifiable}: ship conformance profiles and black-box tests that validate the invariants in
  Table~\ref{tab:invariants} without requiring privileged database access.
\end{itemize}
The reference runtime uses in-memory implementations for the app/key store, audit sink, and rate limiter to keep the
codebase minimal.
Production deployments SHOULD replace these with durable stores and distributed admission control, while preserving the
observable protocol semantics.
The minimal runtime also intentionally omits some optional controls (for example, step-up verification and advanced quota
dimensions); these can be introduced as additional profiles without changing the core protocol envelope.

\subsection{Gateway Architecture and Component Decomposition}
The reference runtime is organized as a gateway and a small admin plane.
The HTTP layer performs schema validation and returns standardized envelopes; request handling is delegated to
governance engines that implement authentication, authorization, write control, and audit emission.
The module decomposition is designed to align directly with the invariants and reason codes described earlier.
Table~\ref{tab:ref-components} summarizes the main components and their responsibilities.

\begin{table}[t]
\caption{Reference runtime components (illustrative).}
\label{tab:ref-components}
\centering
\small
% Constrain width for single-column readability.
\begin{floatbox}
{\setlength{\tabcolsep}{4pt}
\renewcommand{\arraystretch}{1.05}
\begin{tabular}{>{\raggedright\arraybackslash}p{0.34\linewidth} >{\raggedright\arraybackslash}p{0.58\linewidth}}
\toprule
\textbf{Component} & \textbf{Responsibility} \\
\midrule
HTTP gateway (\texttt{/api/agent/v1/*}) &
Route registration, schema validation, and standardized success/error envelopes. \\
Auth + admission control &
Bearer token parsing, app/key resolution, revocation and expiry checks, IP allowlists, and fixed-window rate limiting
before any tool logic. \\
Policy engine &
Scope checks and ABAC constraints (resource allowlists, bounded query windows, redaction), plus tenant/workspace
boundary enforcement. \\
Tool registry &
Authorization-dependent discovery output (\texttt{/manifest}) and tool bindings (metadata + input/output schemas +
optional impact computation for high-risk tools). \\
Draft/execution store &
Write intent tracking (drafts), execution results, and idempotent replay mapping by
$(\mcode{appId}, \mcode{idempotencyKey})$. \\
Audit service &
Structured allow/deny/fail events with correlation identifiers (app/key/actor/draft/execution); pluggable sinks. \\
Admin plane (\texttt{/api/agent-admin/v1/*}) &
App/key lifecycle, policy and auto-execute updates, draft approval/rejection, and audit export. \\
\bottomrule
\end{tabular}
}
\end{floatbox}
\end{table}

For simplicity, the reference runtime uses a deliberately minimal admin authentication mechanism.
This is not a recommended security pattern: production deployments MUST protect the admin plane with operator
authentication and authorization, and they SHOULD apply step-up for high-risk approvals.

\subsection{Adapter Boundary and Tool Registry}
OpenPort standardizes governance semantics, not domain semantics.
The reference runtime therefore defines a narrow \emph{domain adapter} boundary that exposes only the minimal reads and
writes needed by tools, and it passes an \texttt{actorUserId} that represents the effective subject (service user or
delegated user).
This boundary is the main extraction guardrail:
product-specific identity models, joins, and business rules remain private, while the gateway enforces uniform
governance around them.
In practice, adopting OpenPort is primarily an integration exercise: implement the adapter boundary for the target
application and register the tool set to expose, while reusing the protocol-level enforcement for scopes, policies,
drafts, audits, and rate limits.

Tools are registered in a tool registry that produces the authorization-dependent manifest (Eq.~\eqref{eq:visible-tools})
and resolves action implementations for \texttt{POST /api/agent/v1/actions}.
Each tool carries:
\begin{itemize}
  \item \textbf{governance metadata}: \texttt{requiredScopes}, risk tier, and whether confirmation is required;
  \item \textbf{machine-readable schemas}: input/output schemas for client-side validation and tracing;
  \item \textbf{execution binding hooks}: for high-risk tools, an optional impact computation used by
  \texttt{/preflight} and by Eq.~\eqref{eq:preflight-hash}.
\end{itemize}
This structure makes risk controls explicit: risk tier affects whether preflight binding and idempotency are required
for execution, and it determines whether auto-execution can be enabled at all.

\subsection{Conformance Kit and Executable Profiles}
To keep the protocol testable across deployments, the repository includes a machine-readable conformance profile and an
executable runner.
A profile declares:
(i) required endpoints (method + path),
(ii) envelope requirements for success and error cases, and
(iii) minimum security expectations (for example, unauthenticated discovery is denied).
The runner can operate in two modes: an in-process mode against the reference runtime (HTTP injection), and a remote
mode that validates a deployment using only HTTP requests and a provided agent token.

The current core profile focuses on safe, domain-neutral checks:
it validates \texttt{/manifest}, a small read surface, preflight behavior for a high-risk tool, and draft retrieval.
Because OpenPort defaults writes to drafts, conformance can exercise the write pipeline without requiring destructive
side effects, which makes it feasible to run profiles in public CI.
As deployments mature, OpenPort SHOULD be validated using layered profiles (core/authz/writes/abuse/admin) as described
in Section~\ref{sec:invariants}.

\subsection{Release Gates for Safe Open-Source Evolution}
OpenPort is intended to be extracted from private systems without leaking secrets or private implementation details.
The reference repository therefore treats release engineering as part of the governance story: protocol correctness is
not sufficient if the open-source distribution can accidentally publish credentials, internal hostnames, or product
markers.

The release gate pipeline operationalizes this constraint by combining:
\begin{itemize}
  \item \textbf{Build and unit tests}: type checking and regression tests for security controls (scope/policy boundaries,
  redaction, draft approvals, immediate revocation).
  \item \textbf{Conformance profile execution}: a black-box runner that validates endpoint presence and baseline
  semantics.
  \item \textbf{Safety scans}: checks for accidental environment files, common secret patterns (keys, private key
  blocks, tokens), and private-environment markers.
  \item \textbf{Boundary leak scans}: project-specific keyword checks intended to catch accidental references to private
  codebases or infrastructure in public assets.
\end{itemize}
These gates reduce the risk that protocol iteration silently expands the agent authority surface or leaks sensitive
material during open-source publication.

\subsection{Fuzz/Abuse Regression and Negative Security Tests}
Finally, the reference repository includes fuzz-style and abuse-oriented regression tests that target production failure
modes rather than happy-path demos.
The test suite checks that malformed-but-valid JSON inputs do not produce 5xx regressions, that OpenAPI contracts match
the reference server routes, and that key security controls are enforced:
cross-tenant boundary denials, IP allowlists, bounded query windows, draft approval flows, and unit-tested admission
control logic (fixed-window limiting).
Together with conformance profiles, these tests provide executable evidence that OpenPort's governance semantics remain
stable across releases, which is a prerequisite for multi-model agent runtimes to implement safe client behavior.

\section{Validation and Preliminary Evaluation}
\label{sec:evaluation}
OpenPort is a governance specification. We therefore evaluate it as a set of externally observable invariants rather
than as a task-level benchmark (``success rate'' or ``fewer tokens'').
The primary question is whether an agent runtime can converge to safe behavior under realistic production failure modes:
insufficient scopes, policy denials, revocation, retries, prompt-injection-driven tool misuse, and overload.

\subsection{Methodology: Artifact-Based Validation}
We combine three validation artifacts that can be run without privileged access to a product database:
\begin{itemize}
  \item \textbf{Black-box conformance}: a profile-driven harness that interacts only through protocol endpoints and
  validates invariants (Table~\ref{tab:invariants}).
  \item \textbf{Negative security tests}: regression tests for cross-tenant boundaries, policy windows, redaction, key
  revocation, and approval flows.
  \item \textbf{Fuzz/abuse regression}: malformed inputs and abuse-oriented regressions to ensure predictable error
  behavior (no 5xx regressions; stable envelopes and machine-readable codes on 4xx paths).
\end{itemize}
This methodology mirrors real adoption: third-party integrators and agent runtimes observe only the gateway interface
and reason codes, not the application's internal schema.

\subsection{Experimental Setup}
We run the validation suite against a reference deployment consisting of:
\begin{itemize}
  \item an OpenPort gateway configured with a small, deterministic tool registry
  \item a domain adapter seeded with synthetic multi-tenant data (multiple ledgers/workspaces)
  \item multiple integration apps/keys to cover distinct authorization configurations:
  \begin{itemize}
    \item a workspace-scoped app restricted to a single organization
    \item a personal-scoped app with narrower resource allowlists
    \item at least one key under rotation and one revoked key
  \end{itemize}
\end{itemize}
Policies are configured to exercise ABAC constraints (allowed resource IDs, bounded query windows as in
Eq.~\eqref{eq:max-days}, and sensitive-field redaction as in Eq.~\eqref{eq:present}) and network constraints (IP
allowlists).

\subsection{Metrics}
We report outcomes as protocol-level properties rather than task-level agent benchmarks:
\begin{itemize}
  \item \textbf{Artifact pass/fail}: whether build, unit tests, conformance profiles, and safety scans pass at a pinned
  tag.
  \item \textbf{Invariant coverage}: which protocol invariants (Table~\ref{tab:invariants}) are exercised by the
  validation artifacts at that tag.
  \item \textbf{Robustness budget}: whether malformed inputs and negative paths avoid 5xx regressions and preserve
  stable response envelopes.
\end{itemize}
Expanded profiles can additionally report protocol-level metrics such as reason-code stability (Eq.~\eqref{eq:first-fail}),
end-to-end 429 semantics (Eq.~\eqref{eq:fixed-window}), audit completeness (Table~\ref{tab:audit-fields}), and idempotent
replay behavior (Eq.~\eqref{eq:idempotency-map}).

\subsection{Artifacts and Reproducibility}
To keep the evaluation maintainable while the project evolves, we pin artifact-based claims to a tagged release.
At \texttt{v0.1.0}, the reference repository includes a release gate script that executes build, unit tests, a core
conformance profile in local mode, and safety/boundary-leak scans.
The recommended entry point is:
\begin{lstlisting}[language=bash,caption={Reproducible validation entry point (tag \texttt{v0.1.0}).}]
npm run gate
\end{lstlisting}
The conformance runner can also be executed independently (local or remote mode), allowing third-party deployments to
publish a profile conformance claim without sharing internal schemas.
Scripts and profile definitions may evolve; the \texttt{v0.1.0} tag provides a stable reference point for the claims in
this paper.
Optional stronger profiles (for example, State Witness / Preconditions) are treated as extensions and are evaluated separately
from the pinned \texttt{v0.1.0} claims.

\subsection{Coverage at Tag \texttt{v0.1.0}}
Table~\ref{tab:eval-coverage} summarizes how the \texttt{v0.1.0} artifacts map to the protocol invariants.
The intent is not to claim that all OpenPort requirements are fully validated by the minimal suite, but to make the
validation surface explicit and externally reproducible.
We label a property as \emph{Covered} when a runnable artifact includes an automated regression that fails if the
property is violated, \emph{Partial} when validation exists only at the component level (for example, unit-tested admission
control without an end-to-end 429 profile), and \emph{Future} when the requirement is specified but not yet exercised by
the minimal suite.

\begin{table}[t]
\caption{Evaluation coverage at tag \texttt{v0.1.0} (representative).}
\label{tab:eval-coverage}
\centering
\small
% Constrain width for single-column readability.
\begin{floatbox}
{\setlength{\tabcolsep}{4pt}
\renewcommand{\arraystretch}{1.05}
\begin{tabular}{>{\raggedright\arraybackslash}p{0.34\linewidth} >{\raggedright\arraybackslash}p{0.48\linewidth} >{\raggedright\arraybackslash}p{0.12\linewidth}}
\toprule
\textbf{Property / invariant} & \textbf{Artifact(s)} & \textbf{Status} \\
\midrule
Envelope stability &
Core conformance runner; OpenAPI contract test & Covered \\
Unauthenticated discovery denied &
Core conformance unauthorized \texttt{/manifest} & Covered \\
Policy boundaries (tenant/IP/max-days) &
Security controls tests; policy tests & Covered \\
Draft-first pipeline + approval &
App tests (draft + admin approve); conformance draft retrieval & Covered \\
Preflight hash produced &
Conformance \texttt{/preflight}; app high-risk flow & Covered \\
State witness preconditions &
Optional profile; execution-time revalidation regression planned & Future \\
Idempotency replay mapping &
Specified; no dedicated replay regression yet & Future \\
Rate limiting end-to-end 429 &
Limiter unit test; extended profile recommended & Partial \\
Audit completeness (allow/deny/fail) &
Schema and admin listing exist; coverage expansion planned & Future \\
\bottomrule
\end{tabular}
}
\end{floatbox}
\end{table}

\subsection{Preliminary Results (Tag \texttt{v0.1.0})}
At \texttt{v0.1.0}, the release gate passes end-to-end on Node $\ge 20$:
\begin{itemize}
  \item TypeScript build completes without errors.
  \item Unit tests pass (8 test files; 20 tests), covering policy boundaries, draft approval flows, immediate revocation,
  contract-to-route consistency, and fuzz regressions (80 malformed requests across action and query surfaces; no 5xx).
  \item The core conformance profile passes in local mode, validating the presence of 6 required endpoints, envelope
  stability, preflight behavior for a high-risk tool, and draft retrieval.
  \item The tag asserts 3 distinct stable reason codes in unit tests (\url{agent.token_invalid}, \url{agent.policy_denied},
  \url{agent.idempotency_required}); expanding reason-code stability regressions is future work.
  \item Abuse-oriented regressions include a unit test of the fixed-window limiter; an end-to-end 429 ``no drafts on denial''
  profile is future work.
  \item Safety and boundary-leak scans detect no accidental \texttt{.env} files, no common secret patterns, and no
  forbidden private markers in public assets.
\end{itemize}
These results do not claim production readiness of the minimal runtime; they demonstrate that the protocol's governance
semantics are testable, executable, and regression-protected in an open-source workflow.

\section{Related Work}
OpenPort Protocol sits at the intersection of tool exposure for agent runtimes and production API security governance.
Adjacent work typically provides either (i) machine-readable interface contracts and invocation mechanics, or (ii)
delegated identity and token transport.
Neither class typically specifies the closed-loop governance semantics required for safe agent tool exposure:
deny-by-default discovery, least-privilege scopes plus ABAC constraints, risk-gated writes that fail closed into drafts,
stable reason codes, admission control that is side-effect free, and mandatory auditability.
OpenPort makes these behaviors part of the protocol surface and pairs them with externally verifiable invariants
(Section~\ref{sec:invariants}).

\subsection{Tool Exposure and Schema Contracts}
Interface contracts are frequently described using OpenAPI~\cite{openapi}, which standardizes endpoint and schema
descriptions.
Agent tooling ecosystems similarly rely on machine-readable tool catalogs and input/output schemas to avoid brittle UI
automation.
These approaches address \emph{what} tools exist and \emph{how} to call them, but they often leave \emph{when} a tool is
visible, \emph{how} denials are represented, and \emph{how} writes are governed as implementation-defined.
OpenPort is compatible with contract descriptions and tool registries, but it adds governance semantics that are
critical under agent failure modes: authorization-dependent discovery (Eq.~\eqref{eq:visible-tools}), stable
\texttt{agent.*} denial codes, and a draft-first write path that prevents side effects by default.

\subsection{Agent Tool Exposure Protocols and Registries}
Recent agent ecosystems standardize tool discovery and invocation by treating applications as ``tool servers'' and by
exposing machine-readable tool metadata (names, descriptions, schemas) that a client runtime can enumerate and call.
This direction reduces reliance on UI automation and improves interoperability across models and runtimes.
Model Context Protocol (MCP), for example, defines a server that exposes tools and schemas and a client runtime that can
discover and invoke them~\cite{mcp_intro,mcp_spec}.
In the browser, WebMCP proposes a Web API (\texttt{window.navigator.modelContext}) for exposing JavaScript-implemented
tools within a loaded page, explicitly framing such pages as a Web binding of MCP-style tool servers~\cite{webmcp_readme,webmcp_proposal}.
However, tool exposure protocols often leave authorization closure, write governance, and operable failure semantics to
each server implementation: which tools are visible under which credentials, how retries are guided under denial codes,
how high-risk writes are gated behind review, and which invariants are externally verifiable.
OpenPort positions governance as the missing standardization layer: it can bind to existing tool catalogs, but it
requires the closed-loop semantics that make tool exposure safe under agent failure modes (draft-first writes, stable
reason codes, admission control without side effects, and mandatory auditability).

\subsection{Delegated Authorization, Revocation, and Token Binding}
Delegated authorization commonly relies on OAuth 2.0~\cite{rfc6749} and bearer token usage~\cite{rfc6750}, with optional
token introspection~\cite{rfc7662} and revocation~\cite{rfc7009}.
These standards define identity delegation and token lifecycle primitives, but they do not mandate the governance
semantics needed for agent tooling: least-privilege scope mapping, policy windows, risk-gated writes, and audit
completeness.
OpenPort therefore treats delegated identity as an input to a server-side mapping that yields minimal scopes and policy
constraints, and it requires that revocation is immediately effective across discovery and tool calls with stable denial
semantics.

Token theft is a central production risk. Proof-of-possession variants such as mutual TLS bound access tokens~\cite{rfc8705}
and DPoP~\cite{rfc9449} reduce replay by binding a token to a key.
OpenPort's minimal runtime uses bearer tokens for simplicity, but the protocol is compatible with PoP-style bindings:
PoP strengthens authentication, while OpenPort focuses on authorization closure and safe write semantics even when a
token is valid-looking.

\subsection{Policy and Data-Domain Restriction}
Classical access-control models such as RBAC~\cite{sandhu1996} and ABAC-style policy constraints~\cite{nist800162} provide
flexible mechanisms for restricting the data domain and presentation surface beyond coarse scopes.
OpenPort uses policy to encode resource allowlists, bounded query windows (Eq.~\eqref{eq:max-days}), and deterministic
redaction (Eq.~\eqref{eq:present}).
Unlike generic policy frameworks, OpenPort couples these constraints to the tool exposure surface itself: discovery,
reads, drafts, and executions are all governed by the same enforcement posture and stable denial taxonomy.

\subsection{Operational API Governance: Admission Control and Auditable Recovery}
Operational patterns such as admission control and predictable 429 semantics~\cite{rfc9110,rfc6585} are well understood
for conventional APIs, but agent runtimes amplify their importance because retries, polling, and partial failures are
common.
OpenPort specifies rate limiting as a protocol invariant: rate-limited requests must fail before allocating drafts or
side effects, and they must return stable, machine-actionable codes that drive deterministic client behavior
(Table~\ref{tab:recovery}).
Auditability is similarly first-class: structured allow/deny/fail events and stable reason codes enable incident
response and make governance properties testable as regressions rather than as informal best practices.

\subsection{Write Preconditions and TOCTOU}
Many APIs mitigate lost updates and time-of-check to time-of-use (TOCTOU) hazards using preconditions and optimistic
concurrency patterns (for example, \texttt{ETag}-style validators and conditional requests)~\cite{rfc9110}.
These mechanisms are typically scoped to a resource representation and are applied at the time of mutation.
OpenPort adapts the same idea to agent tool exposure: high-risk writes may be subject to delayed operator approval, so
the relevant ``world state'' can change between preflight and execution.
The optional State Witness / Preconditions profile therefore binds a write intent to a server-observed witness hash
(Eq.~\eqref{eq:state-witness-hash}) and requires execute-time revalidation.
On mismatch, the server fails closed with a stable \url{agent.precondition_failed} reason code so clients can rerun
preflight and obtain renewed approval rather than retrying blindly.

\subsection{Capability Attenuation}
Capability-based approaches model authorization as transferable, attenuable tokens.
Macaroons, for example, represent bearer-style capabilities with contextual caveats that can restrict delegation~\cite{macaroons}.
OpenPort can be viewed as complementary: it expresses attenuation via scopes and ABAC policy constraints at the gateway,
adds a governed write pipeline (draft-first, preflight binding, idempotency), and requires an audit trail so that
authorization decisions are operable in real deployments.

Across these areas, OpenPort does not replace interface description or delegation standards; it constrains their
composition for agent runtimes by defining a uniform recovery and governance surface.
This emphasis on stable denial semantics, risk-gated writes, and verifiable invariants distinguishes OpenPort from
specifications that stop at tool schema exposure or token transport.

\section{Limitations and Future Work}
OpenPort is designed to be a domain-neutral governance interface. That choice implies deliberate constraints.

\subsection{Limitations}
\begin{itemize}
  \item \textbf{Not a prompt-safety solution}: OpenPort does not prevent prompt injection or malicious instruction
  following; it constrains the \emph{effects} of tool calls via least privilege, draft-first writes, and operator review.
  \item \textbf{Trusted gateway and admin plane}: the threat model assumes the operator controls the gateway and the admin
  control plane. OpenPort does not defend against a malicious administrator.
  \item \textbf{Domain correctness is adapter-defined}: the protocol enforces scopes, policies, and write governance, but
  it cannot guarantee that an adapter's business logic is correct or safe; adoption requires careful adapter review and
  internal defense-in-depth.
  \item \textbf{Minimal runtime is not production-hardened}: at \texttt{v0.1.0}, governance state (apps/keys/drafts/
  executions) and audit storage are in-memory by default, and admission control uses a process-local fixed-window limiter.
  Multi-node deployments require durable storage for governance state, a distributed limiter, and consistent idempotency
  behavior across replicas.
  \item \textbf{Bearer tokens in the minimal runtime}: the reference implementation uses bearer tokens for simplicity;
  this does not provide proof-of-possession and therefore does not reduce replay risk under token theft.
  \item \textbf{Risk classification is operator-defined}: OpenPort enforces write safeguards based on tool risk metadata;
  incorrect risk labeling or incomplete impact computation can weaken protections and must be treated as a governance
  responsibility.
  \item \textbf{Audit integrity is out of scope for the minimal profile}: OpenPort mandates structured allow/deny/fail
  events with stable codes, but it does not require tamper-evident audit integrity (signing or hash chaining) in the
  minimal runtime; deployments must secure the audit sink and retention pipeline.
  \item \textbf{Optional controls are not fully realized}: some governance mechanisms (for example, step-up verification and
  richer per-tool quotas) are described as profile extensions but are not enforced by the minimal core profile.
\end{itemize}

\subsection{Future Work}
We view OpenPort profiles as the primary mechanism to evolve the protocol without breaking existing clients. Priority
future work includes:
\begin{itemize}
  \item \textbf{Standard delegated-auth bindings}: a hardened OAuth 2.0 binding profile that specifies claim-to-scope
  mapping, policy windows, and revocation behavior, including safe UX patterns for agent consent.
  \item \textbf{Proof-of-possession modes}: optional PoP bindings (for example, mTLS- or DPoP-style) to reduce replay under token
  theft, while preserving the same authorization and audit semantics.
  \item \textbf{Multi-node governance semantics}: a persistence profile and reference implementation for apps/keys/drafts/
  executions (with migration guidance for systems that already have internal audit and authorization pipelines), plus
  guidance and conformance profiles for consistent idempotency mapping, draft/execution correlation, and rate limiting
  across replicas, including an end-to-end 429 profile that proves ``no drafts on rate-limit denials.''
  \item \textbf{Audit integrity and export}: standardized audit export formats and optional integrity protection (event
  signing or hash chaining) suitable for SIEM ingestion and compliance retention.
  \item \textbf{Expanded conformance profiles}: publish and validate layered profiles for authZ, writes, abuse controls,
  and admin-plane security; expand black-box tests for audit completeness and denial-path side-effect freedom.
  \item \textbf{Cost-aware quotas}: per-tool budgets and query-cost controls to bound expensive exports and large
  retrievals beyond fixed request rate limits.
\end{itemize}

\section{Conclusion}
Safe tool exposure for AI agents is primarily a governance problem: runtimes must be able to discover and invoke tools
without expanding the authority surface, and operators must be able to revoke, rate-limit, and audit those capabilities
under real-world failure and abuse.
This paper introduced OpenPort Protocol, a governance-first specification for exposing application data and actions to
AI agents through a stable, machine-readable tool interface.
OpenPort closes the authorization loop by making tool discovery authorization-dependent, enforcing explicit scopes plus
ABAC policy windows, and standardizing stable response envelopes and \texttt{agent.*} reason codes that enable
deterministic client recovery.
For writes, OpenPort defaults to draft creation and supports risk-gated execution via preflight impact hashing,
idempotency, and separation-of-duties approvals, so that high-risk operations fail closed into reviewable intent rather
than side effects.
Operationally, admission control and auditability are treated as protocol invariants: rate-limited requests must be
side-effect free, and allow/deny/fail paths must emit structured audit events for incident response and compliance.

A governance protocol is only as useful as its verifiability.
By pairing a reference runtime with layered profiles, black-box conformance tests, and release-gate automation pinned to
immutable tags (for example, \texttt{v0.1.0}), OpenPort makes safety properties executable and regression-protected as the
standard evolves.
OpenPort is intentionally model- and runtime-neutral and can bind to existing tool ecosystems while keeping governance
server-side.
Future work focuses on durable multi-node semantics for governance state, stronger token-binding options, and expanded
conformance profiles that validate audit completeness and end-to-end abuse controls.

\section*{Stewardship}
OpenPort Protocol is stewarded by \textbf{Accentrust} and the OpenPort Protocol authors:
\textbf{Genliang Zhu}, \textbf{Chu Wang}, \textbf{Ziyuan Wang}, \textbf{Zhida Li}, and \textbf{Qiang Li}.
The project is governed as an open specification paired with a reference runtime and executable conformance artifacts.
Stewardship emphasizes: (i) security-first defaults over convenience-first defaults, (ii) stable, machine-actionable
semantics (envelopes and reason codes) for long-lived agent runtimes, and (iii) verifiable profiles that prevent silent
expansion of the agent authority surface.

\subsection*{Versioning and Compatibility}
OpenPort uses explicit versioning in its URL path (for example, \texttt{agent/v1}).
The reference runtime is released under Semantic Versioning with immutable annotated tags (for example, \texttt{v0.1.0}) used to
pin evaluation claims (Section~\ref{sec:evaluation}).
Within a major version, backward compatibility means: stable response envelopes, stable reason-code semantics, and
conservative additive evolution of tool metadata and endpoints.
Changes that alter authorization semantics, reason-code meaning, or safety invariants SHOULD be introduced as a new
profile identifier or a new major version rather than modifying existing conformance claims.

\subsection*{Profiles and Conformance Claims}
Implementations SHOULD publish which OpenPort profiles they satisfy (core/authz/writes/abuse/admin).
A conformance claim is meaningful only if it is executable: profiles should be machine-readable, runnable against a
deployment in black-box mode, and stable over time.
This paper treats conformance and release gates as part of the protocol: they are the mechanism that allows an open
governance standard to evolve safely.

\subsection*{Specification Change Control}
OpenPort is intended to be implemented by multiple agent runtimes and multiple applications; silent behavioral drift is
therefore treated as a standards failure.
Protocol changes that affect client recovery (response envelopes, reason codes, retry semantics) or write governance
(draft lifecycle, approval/execution semantics) SHOULD be accompanied by conformance updates and negative tests that fail
when invariants are violated.
Breaking changes SHOULD be introduced via a major version bump or a new profile identifier rather than by changing the
meaning of existing reason codes.
Changes are proposed via public pull requests and are expected to include documentation updates and pass release gates
(build, tests, conformance profiles, and safety scans) before being tagged as a stable release.

\subsection*{Open-Source Safety Boundary}
OpenPort is intended to be extracted from private systems without leaking secrets, identifiers, or product-private
logic.
The reference repository operationalizes this boundary through release gates that scan for common secret patterns and
private markers, and through a design rule that pushes domain-specific behavior behind adapter interfaces rather than
embedding business schemas in core modules.
This boundary is a stewardship requirement: open-source publication must not expand the authority surface or disclose
private implementation details.

\subsection*{Security Disclosures}
Because OpenPort is intended for production tool exposure, vulnerabilities and unsafe-by-default behaviors are treated
as protocol issues.
Security reports SHOULD be submitted privately to the maintainers rather than through public issue trackers, and should
include impact, affected endpoints, and reproducible steps.
Before \texttt{v1.0.0}, only the latest minor release is supported for fixes; issues are addressed by rolling forward via
patch releases and immutable tags.
Stewardship prioritizes rapid triage, reason-code and envelope stability for clients, and regression protection through
release gates, negative security tests, and conformance profiles.
Maintainers aim to acknowledge reports within three business days and provide a severity/scope decision within seven
business days.

\bibliographystyle{IEEEtran}
\bibliography{references}

@misc{openapi,
  author       = {{OpenAPI Initiative}},
  title        = {OpenAPI Specification},
  howpublished = {\url{https://spec.openapis.org/oas/v3.2.0.html}},
  year         = {2025},
  month        = {Sep},
  note         = {Version 3.2.0. Accessed: 2026-02-16}
}

@misc{mcp_intro,
  author       = {{Model Context Protocol}},
  title        = {What is the Model Context Protocol (MCP)?},
  howpublished = {\url{https://modelcontextprotocol.io/docs/getting-started/intro}},
  year         = {2026},
  month        = {Feb},
  note         = {Accessed: 2026-02-17}
}

@misc{mcp_spec,
  author       = {{Model Context Protocol}},
  title        = {Model Context Protocol (MCP) Specification Repository},
  howpublished = {\url{https://github.com/modelcontextprotocol/specification}},
  year         = {2026},
  month        = {Feb},
  note         = {Commit \texttt{90c550bf4a261f1a410d624686e8619403e4dbc4} (main). Accessed: 2026-02-17}
}

@misc{webmcp_readme,
  author       = {Walderman, Brandon and Lee, Leo and Nolan, Andrew and Bokan, David and Sagar, Khushal and Van Opstal, Hannah},
  title        = {WebMCP: Enabling Web Apps to Provide JavaScript-Based Tools},
  howpublished = {\url{https://raw.githubusercontent.com/webmachinelearning/webmcp/971aa24aea2afd865ca8607ba79a486fc7429360/README.md}},
  year         = {2025},
  month        = {Aug},
  note         = {First published: 2025-08-13. Accessed: 2026-02-17}
}

@misc{webmcp_proposal,
  author       = {Walderman, Brandon and Nolan, Andrew and Bokan, David and Sagar, Khushal and Van Opstal, Hannah},
  title        = {WebMCP API Proposal},
  howpublished = {\url{https://raw.githubusercontent.com/webmachinelearning/webmcp/971aa24aea2afd865ca8607ba79a486fc7429360/docs/proposal.md}},
  year         = {2025},
  month        = {Aug},
  note         = {Accessed: 2026-02-17}
}

@techreport{rfc2119,
  author      = {Bradner, S.},
  title       = {Key words for use in RFCs to Indicate Requirement Levels},
  institution = {RFC Editor},
  type        = {RFC},
  number      = {2119},
  year        = {1997},
  url         = {https://www.rfc-editor.org/rfc/rfc2119}
}

@techreport{rfc8174,
  author      = {Leiba, B.},
  title       = {Ambiguity of Uppercase vs Lowercase in RFC 2119 Key Words},
  institution = {RFC Editor},
  type        = {RFC},
  number      = {8174},
  year        = {2017},
  url         = {https://www.rfc-editor.org/rfc/rfc8174}
}

@techreport{rfc6749,
  author      = {Hardt, D.},
  title       = {The OAuth 2.0 Authorization Framework},
  institution = {RFC Editor},
  type        = {RFC},
  number      = {6749},
  year        = {2012},
  url         = {https://www.rfc-editor.org/rfc/rfc6749}
}

@techreport{rfc6750,
  author      = {Jones, M. and Hardt, D.},
  title       = {The OAuth 2.0 Authorization Framework: Bearer Token Usage},
  institution = {RFC Editor},
  type        = {RFC},
  number      = {6750},
  year        = {2012},
  url         = {https://www.rfc-editor.org/rfc/rfc6750}
}

@techreport{rfc7009,
  author      = {Lodderstedt, T. and Dronia, S. and Scurtescu, M.},
  title       = {OAuth 2.0 Token Revocation},
  institution = {RFC Editor},
  type        = {RFC},
  number      = {7009},
  year        = {2013},
  url         = {https://www.rfc-editor.org/rfc/rfc7009}
}

@techreport{rfc9110,
  author      = {Fielding, R. and Nottingham, M. and Reschke, J.},
  title       = {HTTP Semantics},
  institution = {RFC Editor},
  type        = {RFC},
  number      = {9110},
  year        = {2022},
  doi         = {10.17487/RFC9110},
  url         = {https://www.rfc-editor.org/rfc/rfc9110}
}

@techreport{rfc6585,
  author      = {Nottingham, M. and Fielding, R.},
  title       = {Additional HTTP Status Codes},
  institution = {RFC Editor},
  type        = {RFC},
  number      = {6585},
  year        = {2012},
  url         = {https://www.rfc-editor.org/rfc/rfc6585}
}

@techreport{nist800162,
  author      = {Hu, Vincent C. and Ferraiolo, David and Kuhn, Rick and Schnitzer, Arthur and Sandlin, Kenneth and Miller, Robert and Scarfone, Karen},
  title       = {Guide to Attribute Based Access Control (ABAC) Definition and Considerations},
  institution = {National Institute of Standards and Technology},
  type        = {NIST Special Publication},
  number      = {800-162},
  year        = {2014},
  month       = {Jan},
  doi         = {10.6028/NIST.SP.800-162},
  url         = {https://csrc.nist.gov/publications/detail/sp/800-162/final}
}

@techreport{rfc7662,
  author      = {Richer, J.},
  title       = {OAuth 2.0 Token Introspection},
  institution = {RFC Editor},
  type        = {RFC},
  number      = {7662},
  year        = {2015},
  url         = {https://www.rfc-editor.org/rfc/rfc7662}
}

@techreport{rfc8705,
  author      = {Campbell, B. and Bradley, J. and Sakimura, N.},
  title       = {OAuth 2.0 Mutual-TLS Client Authentication and Certificate-Bound Access Tokens},
  institution = {RFC Editor},
  type        = {RFC},
  number      = {8705},
  year        = {2020},
  url         = {https://www.rfc-editor.org/rfc/rfc8705}
}

@techreport{rfc9449,
  author      = {Fett, D. and Jones, M. and Bradley, J.},
  title       = {OAuth 2.0 Demonstrating Proof-of-Possession at the Application Layer (DPoP)},
  institution = {RFC Editor},
  type        = {RFC},
  number      = {9449},
  year        = {2023},
  url         = {https://www.rfc-editor.org/rfc/rfc9449}
}

@techreport{rfc8785,
  author      = {Rundgren, A. and Jordan, B. and Balfanz, D. and Nystrom, M.},
  title       = {JSON Canonicalization Scheme (JCS)},
  institution = {RFC Editor},
  type        = {RFC},
  number      = {8785},
  year        = {2020},
  url         = {https://www.rfc-editor.org/rfc/rfc8785}
}

@techreport{rfc2697,
  author      = {Heinanen, J. and Baker, F. and Weiss, W. and Wroclawski, J.},
  title       = {A Single Rate Three Color Marker},
  institution = {RFC Editor},
  type        = {RFC},
  number      = {2697},
  year        = {1999},
  url         = {https://www.rfc-editor.org/rfc/rfc2697}
}

@techreport{rfc6962,
  author      = {Laurie, B. and Langley, A. and Kasper, E.},
  title       = {Certificate Transparency},
  institution = {RFC Editor},
  type        = {RFC},
  number      = {6962},
  year        = {2013},
  url         = {https://www.rfc-editor.org/rfc/rfc6962}
}

@article{saltzer1975,
  author  = {Saltzer, Jerome H. and Schroeder, Michael D.},
  title   = {The Protection of Information in Computer Systems},
  journal = {Proceedings of the IEEE},
  year    = {1975},
  month   = {Sep},
  volume  = {63},
  number  = {9},
  pages   = {1278--1308},
  doi     = {10.1109/PROC.1975.9939},
  url     = {https://doi.org/10.1109/PROC.1975.9939}
}

@article{sandhu1996,
  author  = {Sandhu, Ravi S. and Coyne, Edward J. and Feinstein, Hal L. and Youman, Charles E.},
  title   = {Role-Based Access Control Models},
  journal = {Computer},
  year    = {1996},
  month   = {Feb},
  volume  = {29},
  number  = {2},
  pages   = {38--47},
  doi     = {10.1109/2.485845},
  url     = {https://doi.org/10.1109/2.485845}
}

@inproceedings{macaroons,
  author    = {Birgisson, Arnar and Politz, Joe Gibbs and Erlingsson, Ulfar and Taly, Ankur and Vrable, Michael and Lentczner, Mark},
  title     = {Macaroons: Cookies with Contextual Caveats for Decentralized Authorization in the Cloud},
  booktitle = {Proceedings of the Network and Distributed System Security Symposium (NDSS)},
  year      = {2014},
  month     = {Feb},
  doi       = {10.14722/ndss.2014.23212},
  url       = {https://www.ndss-symposium.org/ndss2014/macaroons-cookies-contextual-caveats-decentralized-authorization-cloud/}
}

\end{document}